\documentclass[aps,prb,preprint,superscriptaddress,showpacs,amsmath,amssymb,floatfix,tighten]{revtex4-1}
\usepackage{graphicx}
\usepackage{bm}

\usepackage{color}
\usepackage[table]{xcolor}
\usepackage{lipsum,multirow}
\usepackage{amssymb,amsmath,epsfig,rawfonts}
\usepackage{float}
\usepackage{ragged2e} 
\usepackage{mathptmx}
\usepackage{caption} 
\usepackage{array}
\newcolumntype{C}[1]{>{\centering\let\newline\\\arraybackslash\hspace{0pt}}m{#1}} 

\usepackage[breaklinks=true,colorlinks=true,urlcolor=blue,
citecolor=blue,linkcolor=blue,bookmarks=false]{hyperref}
\newcommand{\ltsim}{\protect\raisebox{-0.5ex}{$\:\stackrel{\textstyle <}{\sim}\:$}}

\usepackage{lineno}

\begin{document}
%\begin{linenumbers}

\title{Magnetic mechanism for the biological functioning of hemoglobin}

\author{Selma Mayda}
\affiliation{Department of Physics, Izmir Institute of Technology, 
Urla 35430, Turkey}
\affiliation{Department of Materials Science and Engineering, 
Izmir Institute of Technology, Urla 35430, Turkey}

\author{Zafer Kandemir}
\affiliation{Department of Physics, Izmir Institute of Technology, 
Urla 35430, Turkey}

\author{Nejat Bulut}
\email{nejatbulut@iyte.edu.tr}
\affiliation{Department of Physics, Izmir Institute of Technology, 
Urla 35430, Turkey}

\author{Sadamichi Maekawa}
\affiliation{RIKEN Center for Emergent Matter Science,
Wako 351-0198, Japan}
\affiliation{Kavli Institute for Theoretical Sciences, 
University of Chinese Academy of Sciences, 
Beijing 100049, China}

\date{\today}

\begin{abstract}
The role of magnetism in the biological functioning 
of hemoglobin has been debated
since its discovery by Pauling and Coryell in 1936. 
The hemoglobin molecule contains four heme groups 
each having a porphyrin layer with a Fe ion at the center.
Here, we present combined density-functional theory 
and quantum Monte Carlo calculations
for an effective model of Fe in a heme cluster.
In comparison with these calculations,
we analyze the experimental data on human adult hemoglobin (HbA)
from the magnetic susceptibility, 
M\"ossbauer and 
magnetic circular dichroism (MCD) measurements.
In both the deoxygenated (deoxy) and the oxygenated (oxy) cases, 
we show that local magnetic moments develop 
in the porphyrin layer with antiferromagnetic coupling to the Fe moment.
Our calculations reproduce the magnetic susceptibility measurements 
on deoxy and oxy-HbA. 
For deoxy-HbA,
we show that the anomalous MCD signal
in the UV region is an experimental evidence for the presence 
of antiferromagnetic Fe-porphyrin correlations. 
The various important properties of hemoglobin are explained based 
on the antiferromagnetic correlations 
including the Bohr effect and cooperativity. 
This analysis shows that magnetism is  
involved in a fundamental way in the functioning of hemoglobin.
\end{abstract}

\maketitle

Pauling and Coryell showed that 
the magnetic susceptibility of deoxy-HbA exhibits
a Curie-type ($1/T$) temperature dependence,
while for oxy-HbA it is weakly negative
implying that the total spin $S=0$ for the molecule \cite{Pauling1936,Bren}. 
M\"ossbauer experiments \cite{Lang} are consistent with the view that 
Fe is in an $S=2$ state in deoxy-HbA,
while its magnetic moment is found to be $\ltsim 1\,\mu_{\rm B}$ in the oxy case.
The magnetic circular dichroism (MCD) measurements \cite{Treu}
find an anomalous line shape for the temperature-dependent 
MCD spectra in the UV region.
The HbA molecule exhibits
remarkable functional properties such as the cooperativity 
\cite{Monod,Perutz1970,Yuan}
and the Bohr effect \cite{Adair1925A,Adair1925B,Ferry,Pauling1935}, 
which enhance its oxygen carrying capacity.
When one of the four Fe ions in HbA combines an oxygen molecule, 
the other three Fe ions attract oxygens cooperatively. 
The Bohr effect denotes the characteristic of HbA
through which the oxygen affinity depends on the pH of the medium.
There is currently no broad agreement on 
the spin and charge distributions and the role of magnetism in HbA
\cite{Weiss,Pauling1964,McClure,Goddard,Scheidt,Chen}.

We study the electronic state of one heme group 
by using an effective multi-orbital
Anderson impurity model \cite{Anderson,Haldane}.
The parameters of this model are obtained by
the density functional theory \cite{Kohn} (DFT) calculations,
where we use the molecular coordinates
determined by the X-ray measurements \cite{Perutz1960,Park}.
This way the stereochemical effects are included.
We then study this model with the quantum Monte Carlo (QMC) 
simulations \cite{Hirsch}. 
This DFT+QMC approach is described in the Methods section. 

Figure 1(a) shows the molecular structure of deoxy-HbA \cite{Park}.
We have performed our DFT+QMC calculations for the 
truncated clusters shown in Figs. 1(b) and (c).

Figure 2 shows DFT+QMC results
for the magnetic-moment density $M({\bf r})$ 
at temperature $T=150$ K.
In the deoxy case,
we observe that the Fe site has a large up moment $\approx 4.6\,\mu_{\rm B}$.
The neighboring nitrogen sites have smaller moments,
while at the carbon sites
$M({\bf r})$ points down. 
These down moments originate from the partially-occupied $\pi^*$ host states
consisting of the C($2p_z$) orbitals.
Hence,
antiferromagnetic correlations exist 
between the large Fe magnetic moment 
and the host moments spread out in the porphyrin layer
in deoxy-heme.
In oxy-heme, the Fe moment is reduced, but remains finite.
In addition, the neighboring O$_2$ and N sites
have magnetic moments 
which are antiferromagnetically coupled to that of Fe. 
In the oxy case, the antiferromagnetic screening cloud 
is more tightly localized around the Fe site.
The calculation of $M({\bf r})$ 
is described in the Supplementary Information. 
Additional DFT+QMC data on the spin and charge distributions 
are presented in the Methods section. 

Figure 3(a) shows the temperature dependence 
of the total spin susceptibility $\chi_{\rm t}$.
For the deoxy-heme cluster,
$\chi_{\rm t}$ follows a nearly-perfect Curie $T$-dependence. 
The total effective magnetic moment $M_{\rm t}$
is $\approx 4.1\,\mu_{\rm B}$ at $T=150$ K,
which is reduced from that of Fe 
due to the Fe-porphyrin antiferromagnetic correlations. 
In the oxy case, there are two temperature regimes
separated by a crossover temperature $T^*\approx 300\,{\rm K}$: 
In the high-$T$ regime, $T> T^*$, 
$\chi_{\rm t}$ has a Curie-type $T$ dependence 
with an effective total moment $\approx 2.1\,\mu_{\rm B}$.
In the low-$T$ regime, $T< T^*$, 
$\chi_{\rm t}$ decreases rapidly as $T$ decreases. 
For $T>300$ K,
the reduction of $\chi_{\rm t}$ with respect to that of deoxy-heme 
is mainly due to the collapse of the Fe magnetic moment
because of the loss of the
ferromagnetic correlations among the Fe($3d_{\nu}$) orbitals.
For $T< 300$ K,
$\chi_{\rm t}\rightarrow 0$ with the opening of a magnetic gap 
at the Fermi level.
This is clearly seen in Fig. 3(b),
which shows that the total moment $M_{\rm t}$ gets suppressed 
within $\Delta_{\rm S}\approx 0.15$ eV of the Fermi level as $T$ decreases.
This gap arises from the transfer of electrons from mainly the O$_2$ 
to the Fe($3d_{\nu}$) orbitals, in particular, 
to the so-called $t_{2g}$ orbitals, $\nu=xy$, $xz$ and $yz$.
These are discussed further in the Methods section.

Experimentally, 
the spin susceptibility of oxy-HbA vanishes
at room temperature \cite{Pauling1936,Pauling1977},
while we find that 
$\chi_{\rm t} \approx 150$ $\mu_{\rm B}^2/$eV per heme.
It is possible that because of the various approximations
introduced, the DFT+QMC approach
is underestimating the value of $T^*$. 
If this is indeed the case,
then it will have important consequences for the binding mechanism 
of O$_2$ to Fe in heme and the Bohr effect:
We note that, in general, charge transfer to the Fe($3d$) orbitals 
would be energetically costly 
because of the large Coulomb repulsion over there. 
However, as seen in the Methods section, 
in heme this is overcome by making use of the 
the upper-Hubbard level of the Fe($3d_{xy}$) orbital.
This is where the Fe($3d_{xy}$) orbital becomes doubly occupied,
and it is located very close to the Fermi level. 
This turns out to be an important feature of the 
electronic structure of oxy-heme,
because when the chemical potential is away from this region, 
the magnetic gap does not open. 
We think that, in order to minimize its energy,
the system is developing magnetic moments and magnetic correlations 
by redistributing electrons.
It is a real possibility that the gain in energy 
by going to the $S=0$ state is responsible 
for the binding of O$_2$ to heme. 
These suggest that the binding of O$_2$ to Fe in heme
has a magnetic origin. 
We note that the stereochemical effects \cite{Perutz1970} 
are clearly important in O$_2$ binding. 
We are proposing that the magnetism,
which is controlled by the stereochemical effects,
has the key role in the perfectly reversible binding of O$_2$ to Fe in heme.

In the Bohr effect,
the oxygen affinity of HbA is controlled 
by the hydrogen ion concentration in the red blood cells. 
We have seen above that the $S=0$ state is developing 
within a narrow energy window at the Fermi level. 
Hence,
the magnetic properties depend sensitively on the electron filling. 
Any changes which effectively moves the chemical potential
away from this narrow region will affect the binding of O$_2$. 
This could be
how the pH influences the oxygen affinity. 

We note that 
the above suggested mechanism for the Fe-oxygen bonding 
may not be limited to heme,
but may also be relevant for bonding in other
compounds containing transition metals, for example 
the transition-metal oxides. 
This type of bonding is more complicated than the usual covalent or ionic bondings
because it involves the upper-Hubbard level of the $3d$ orbitals 
and the magnetic correlations. 

An important feature 
which emerges from these calculations 
is the antiferromagnetic coupling between Fe and the porphyrin layer. 
Experimental evidence for this 
is provided by the MCD data on deoxy-HbA in the UV region.
The MCD intensity $\Delta\epsilon(E)$
is the difference between the left-circularly polarized (LCP)
and the right-circularly polarized (RCP) light absorption
within an applied magnetic field parallel to 
the direction of light propagation. 
The MCD spectrum of deoxy-HbA has a peak in the UV region 
at $\approx 3$ eV which has an anomalous line shape 
and a $1/T$ temperature-dependence \cite{Treu}.
The optical absorption $\epsilon(E)$ has a $T$-independent peak 
at the same energy.
It is known that the optical absorption at $\approx 3$ eV is due to 
$\pi \rightarrow \pi^*$ transitions
between the bonding $\pi$ and antibonding $\pi^*$ 
states \cite{Treu}. 
The MCD spectrum of deoxy-HbA in UV region is anomalous in the following sense:
In the usual case, the $T$-dependent piece of the MCD spectrum 
first has a dip and then a peak as the frequency increases,
whereas in deoxy-HbA the MCD spectrum first 
has a peak and then a shallow dip.
Here, we propose that 
the anomalous MCD signal is caused by
the orbital-selective optical transitions from the occupied $\pi$ orbitals
to two partially occupied $\pi^*$ orbitals,
which we label as $\pi_1^*$ and $\pi_2^*$. 
According to the DFT+QMC calculations, 
the $\pi_1^*$ state is nearly half-filled 
while the $\pi_2^*$ is nearly empty,
which is discussed further in the Methods section. 

As illustrated in Fig. 4(a),
within an applied field ${\bf B}_{\rm app}$ pointing up,
the Fe spin will be polarized in the down direction.
Meanwhile,
the spin polarization of $\pi_1^*$ will be parallel to the field,
because of the Fe-$\pi_1^*$ antiferromagnetic correlations. 
Hence, during an LCP (RCP) optical transition from the $\pi$ state,
it will be energetically more favorable for the $\pi_1^*$ ($\pi_2^*$) state 
to absorb a down-spin (up-spin) electron. 
These orbital-selective optical transitions are 
illustrated in Figs. 4(b) and (c). 
We have calculated the MCD spectrum due to these transitions 
as discussed in the Methods section. 
Figures 4(d) and (e) present the experimental and the calculated 
MCD spectra, respectively. 
The inset of Fig. 4(d) illustrates the line shape normally 
expected for the $T$-dependent MCD spectrum \cite{Mason}.
The anomalous MCD line shape of deoxy-HbA seen in Fig. 4(d) 
had been attributed 
to a negative spin-orbit coupling \cite{Treu}.
We are suggesting that it instead
originates from the antiferromagnetic coupling between 
the Fe($3d$) and $\pi^*$ states.
Because of this,
$\pi_1^*$ and $\pi_2^*$ porphyrin states act 
as if they have negative $g$-factors. 
The agreement with the experimental data can be improved further by 
incorporating into the analysis the optical absorption data 
as shown in the Methods section. 

Discussion of cooperativity requires a multi-heme model. 
However, 
because of its importance,
we will briefly comment on the implications 
of the DFT+QMC results. 
Assuming independent heme groups,
the magnetic susceptibility measurements yield an effective magnetic moment
per heme $M_{\rm heme}=5.46 \, \mu_{\rm B}$ 
in deoxy-HbA \cite{Pauling1936}.
Since this is larger than the spin $S=2$ value of 4.9 $\mu_{\rm B}$,
Pauling and Coryell already pointed out the possibility 
of inter-heme ferromagnetic correlations \cite{Pauling1936}. 
The DFT+QMC calculations find $M_{\rm heme}\approx 4.1 \, \mu_{\rm B}$. 
This requires an average inter-heme magnetic correlation 
$\langle M^z_{\rm heme,1} M^z_{\rm heme,2}\rangle 
\approx 4.3\, \mu_{\rm B}^2$ for deoxy-HbA,
which is significant.
We note that 
sufficiently strong inter-heme magnetic correlations
could lead to cooperativity:
The first O$_2$ to bind Fe needs to overcome 
the inter-heme ferromagnetic interaction.
Upon O$_2$ binding to Fe, 
the magnetic moment of that heme group vanishes. 
The following O$_2$'s will bind more easily
because of the loss of the inter-heme ferromagnetic correlations. 
An alternative scenario is 
based on the spin non-conservancy in the binding of O$_2$ to heme. 
We note that 
while O$_2$ is in a triplet state ($S=1$) and deoxy-heme is considered to be $S=2$,
the resultant oxy-heme is in an $S=0$ state. 
This non-conservancy of the total spin in the binding of O$_2$ may be limiting 
the reaction rate. 
However, because of inter-heme ferromagnetic correlations,
spin transfer may be possible among the heme groups.
Hence, when the O$_2$'s bind cooperatively,
the spin can be conserved,
thus eliminating the limit on the O$_2$ binding rate. 

We have shown that magnetic moments and magnetic correlations exist 
in deoxy and oxy-HbA.
In addition, the charge transfer from oxygen to Fe 
and the magnetic gap are key electronic properties of oxy-HbA. 
It is remarkable that evolution has been able to utilize these properties,
which arise from the strongly-correlated electrons,
for the functioning of hemoglobin.
We note that there are a large number of metalloproteins, metalloenzymes 
and other bioinorganic molecules
containing transition-metal centers \cite{Gray}.
Hence,
the magnetic effects are poised to have
a general role in the functioning of 
bioinorganic molecules
with a distinct place in the emerging field of quantum biology. 

%\newpage
\bibliographystyle{apsrev4-1}

\bibliography{mhl2}

\vspace{0.8 cm}
{\bf Acknowledgements}
We thank Nuran Elmaci Irmak, Talat Yalcin, Acar Savaci and Tahir Cagin
for comments. 
Support by the Turkish Scientific and Technical Research Council 
(TUBITAK grant number 113F242) is gratefully acknowledged.
This work was supported by JST ERATO Grant No. JPMJER1402, 
JSPS Grant-in-Aid for Scientific Research on Innovative Areas Grant No. JP26103005, 
and JSPS KAKENHI Grant Nos. JP16H04023 and JP17H02927.
The numerical calculations reported here were performed 
at the TUBITAK ULAKBIM High Performance and Grid
Computing Center (TRUBA resources). 

\vspace{0.8 cm}
%{\bf Additional information}

%{\bf Author contributions}
%. Mayda perfomed the QMC calculations, Z.K. performed the DFT calculations, 
%N.B. and S. Maekawa wrote the paper together.

%\vspace{0.3 cm}
%{\bf Competing interests} The authors declare no competing interests. 

%\vspace{0.3 cm}
%{\bf Supplementary Information} is linked to the online version of the paper at %www.nature.com/nature. 

%\vspace{0.3 cm}
%{\bf Correspondence and requests for materials} should be addressed to N.B. 

%\end{linenumbers}
\clearpage

\begin{figure}
\centering
\caption*{
{\bf FIG. 1 $\vert$ Molecular structure of HbA and the heme clusters}
}
\vspace{1.0cm}
\includegraphics[height=8.0cm]{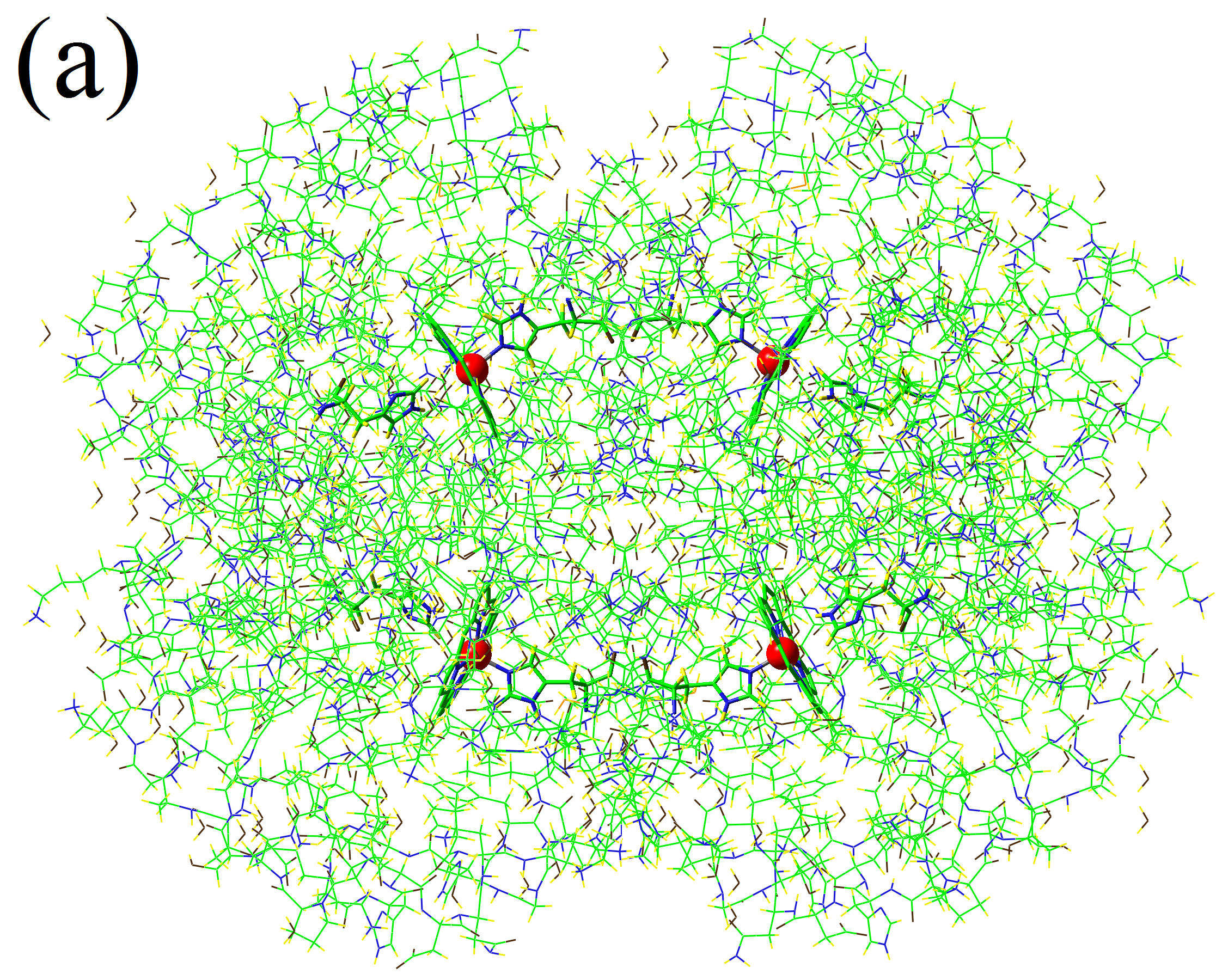}\vspace{1cm}\\
\includegraphics[height=7.5cm]{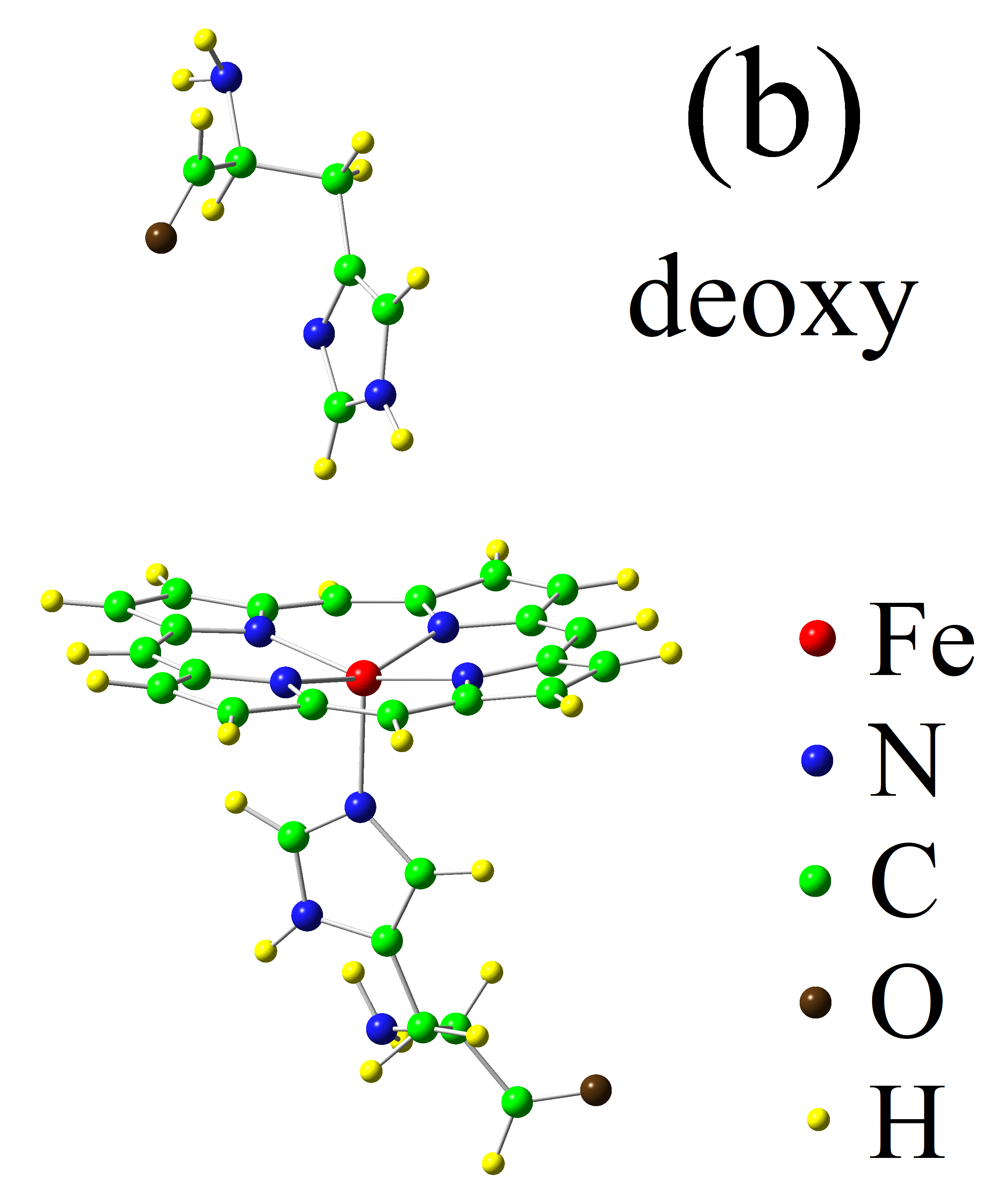}\hspace{2cm}
\includegraphics[height=7.5cm]{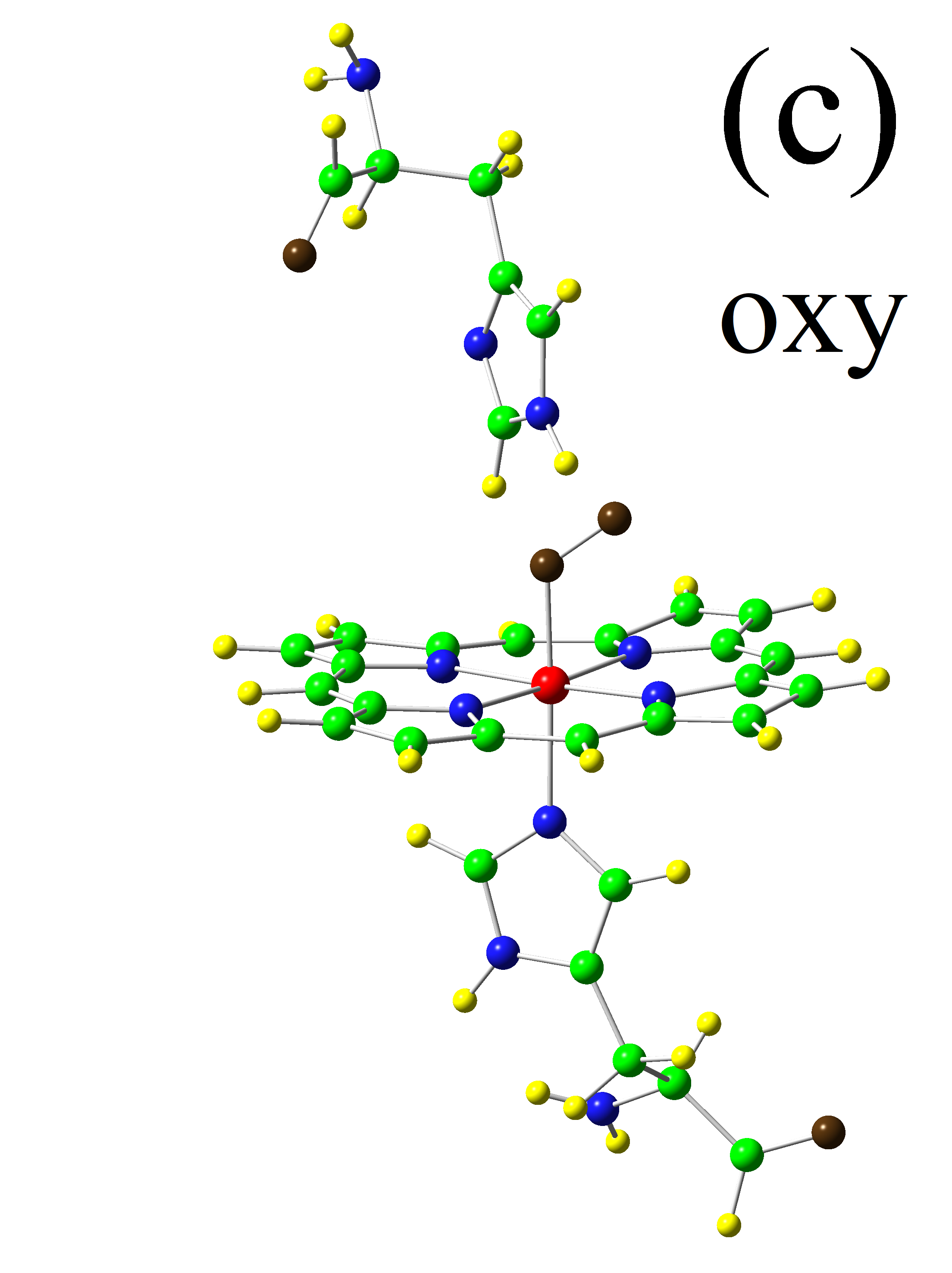}\vspace{1cm}
\justify
(a) Molecular structure of deoxy-HbA
obtained by X-ray diffraction \cite{Park}
(Protein Data Bank, Keyword: 2DN2).
Truncated (b) deoxy and (c) oxy-heme clusters which we have used in the DFT+QMC calculations.
The way these clusters are obtained from the full HbA molecular structures 
is described in the Supplementary Information. 
\label{fig1}
\end{figure}

\begin{figure}
\centering
\caption*{
{\bf FIG. 2 $\vert$ Magnetic-moment density} 
}
\vspace{1.0cm}
\includegraphics[height=9.0cm]{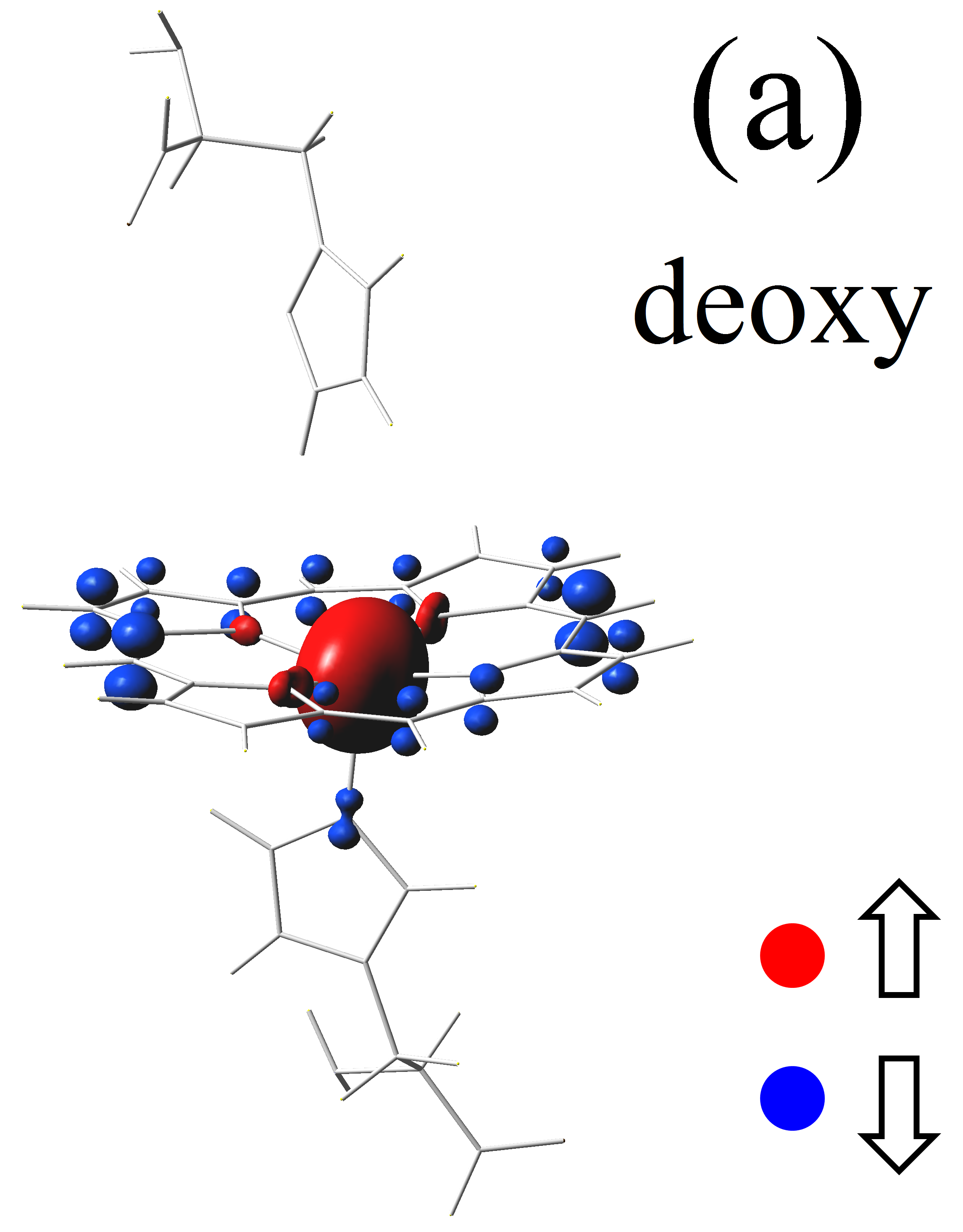}\hspace{2.0cm}
\includegraphics[height=9.0cm]{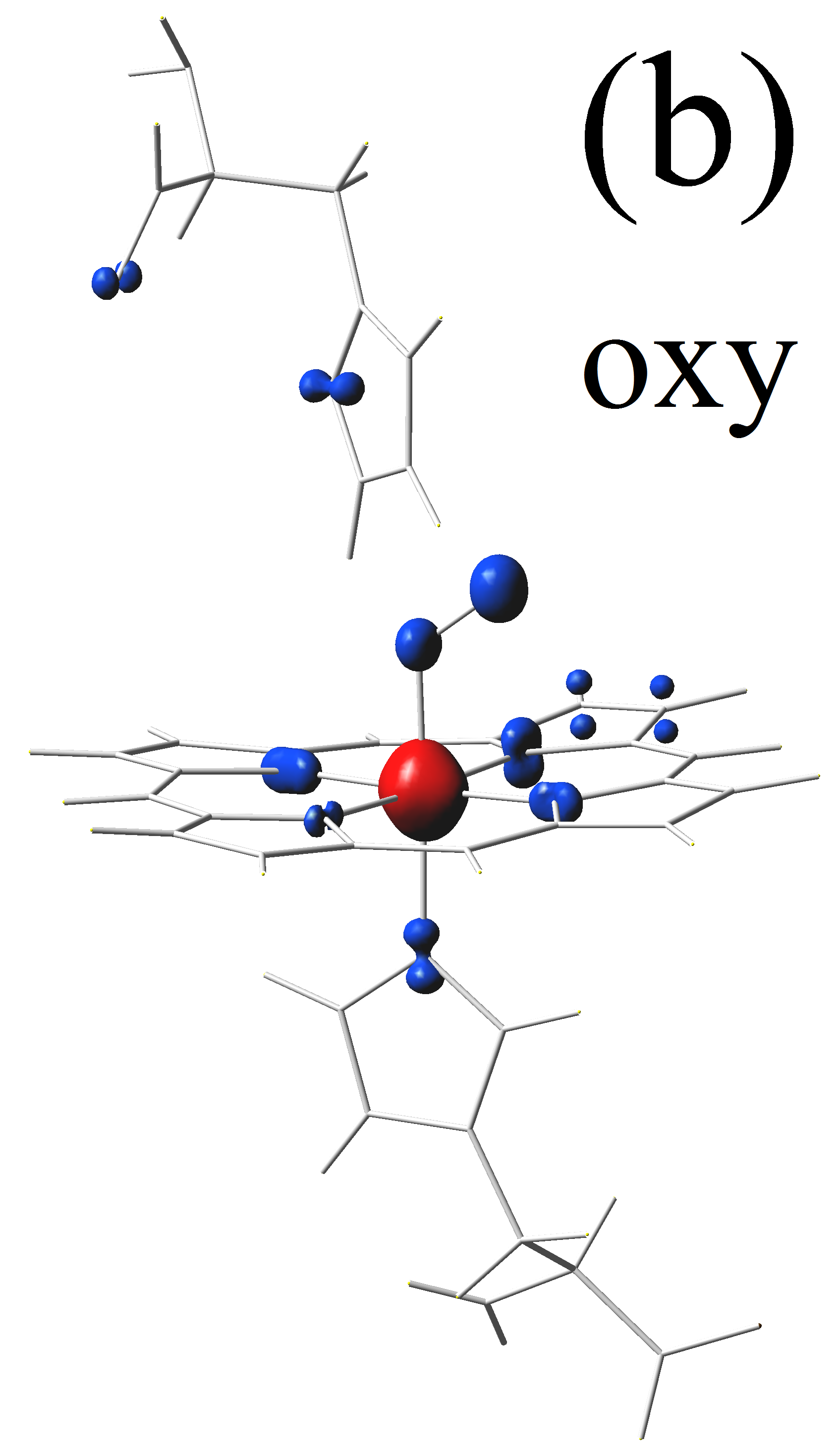}\vspace{1cm}
\justify 
Illustration of the magnetic-moment density $M({\bf r})$ 
for (a) the deoxy and (b) oxy-heme clusters at $T=150$ K.
Here, the red (blue) color indicates the atomic orbitals which have 
magnetic moments pointing up (down). 
The magnitude of $M({\bf r})$ at an atomic site is proportional 
to the volume of the bubble at that site.
The calculation of 
$M({\bf r})$ is described in the Supplementary Information. 
\label{fig2}
\end{figure}

\begin{figure}
\centering
\caption*{
{\bf FIG. 3 $\vert$ Spin susceptibility and the magnetic moment} 
}
\vspace{1.0cm}
\includegraphics[width=7.5cm]{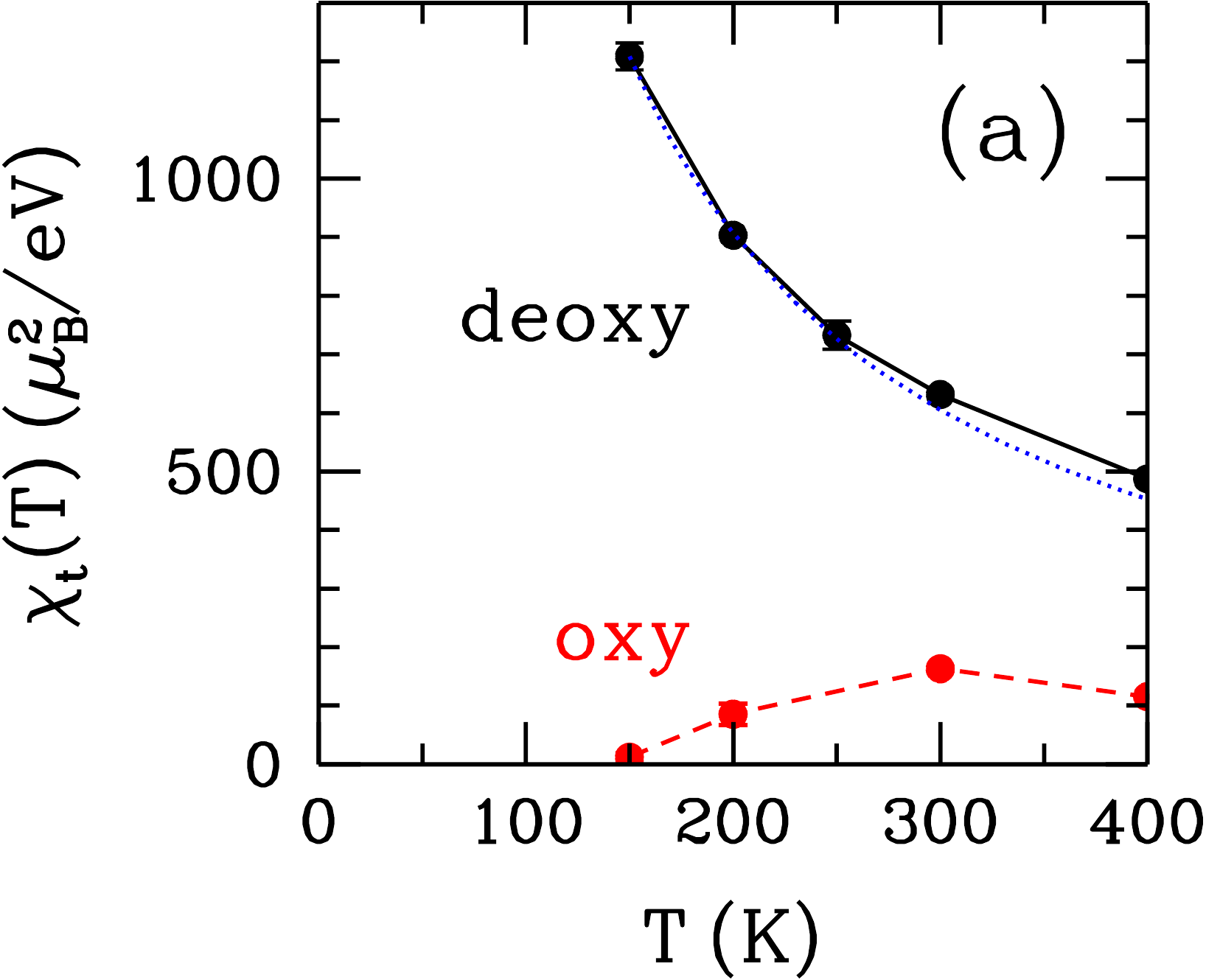}\hspace{1cm}
\includegraphics[width=7.5cm]{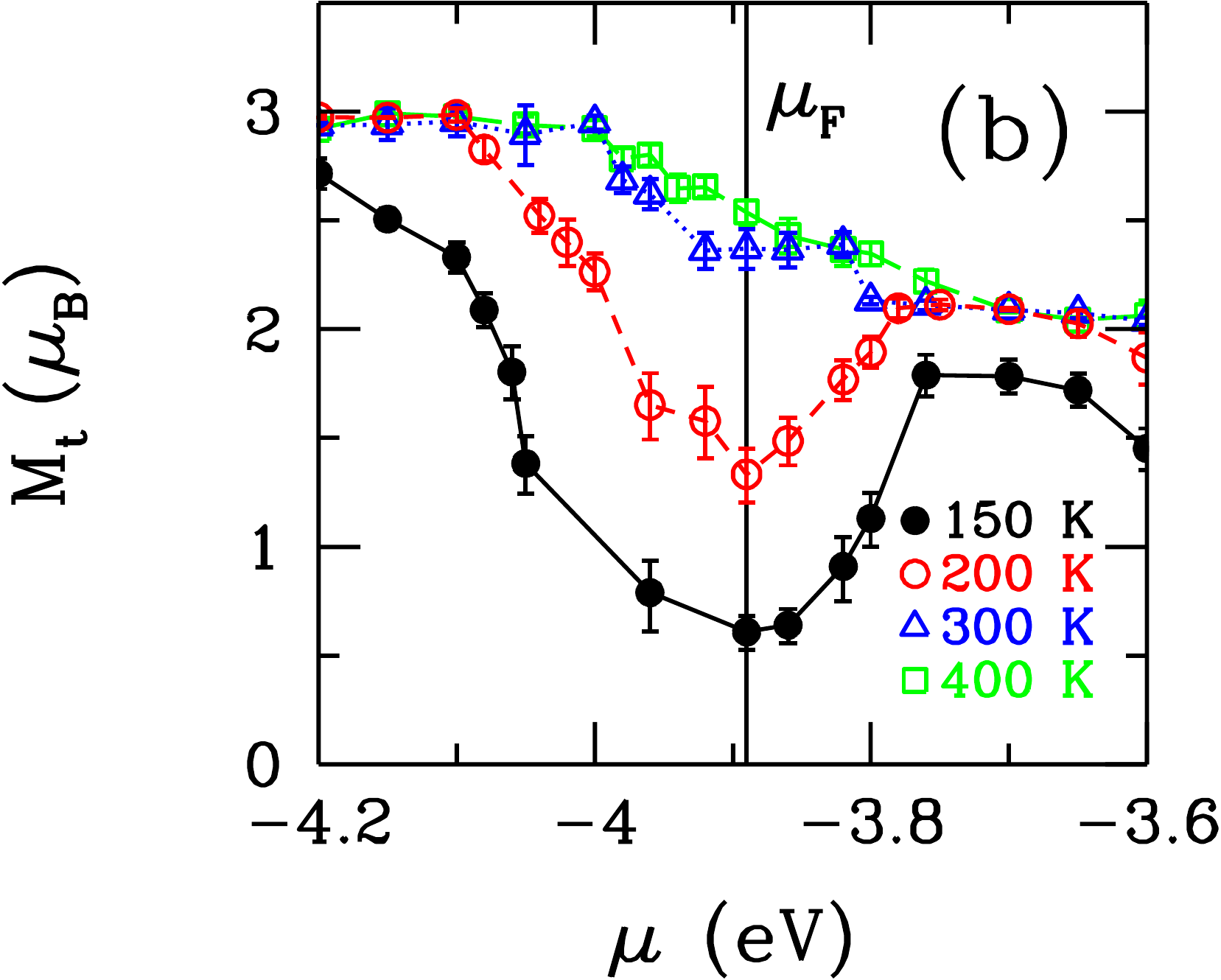}\vspace{1cm}
\justify
(a) Temperature dependence of the total spin susceptibility $\chi_{\rm t}$ for the 
deoxy and oxy-heme clusters. 
The blue dotted curve denotes the Curie ($1/T$) temperature dependence. 
(b) Total magnetic moment $M_{\rm t}$ 
versus the chemical potential $\mu$
near the Fermi level $\mu_{\rm F}$ at various temperatures
for the oxy-heme cluster.
Here,
the black vertical line denotes $\mu_{\rm F}$ at $T=150$ K. 
\label{fig3}
\end{figure}

\begin{figure}
\centering
\caption*{
{\bf FIG. 4 $\vert$ Anomalous MCD spectrum of deoxy-HbA in the UV region} 
}
\vspace{1.0cm}
\includegraphics[height=9cm]{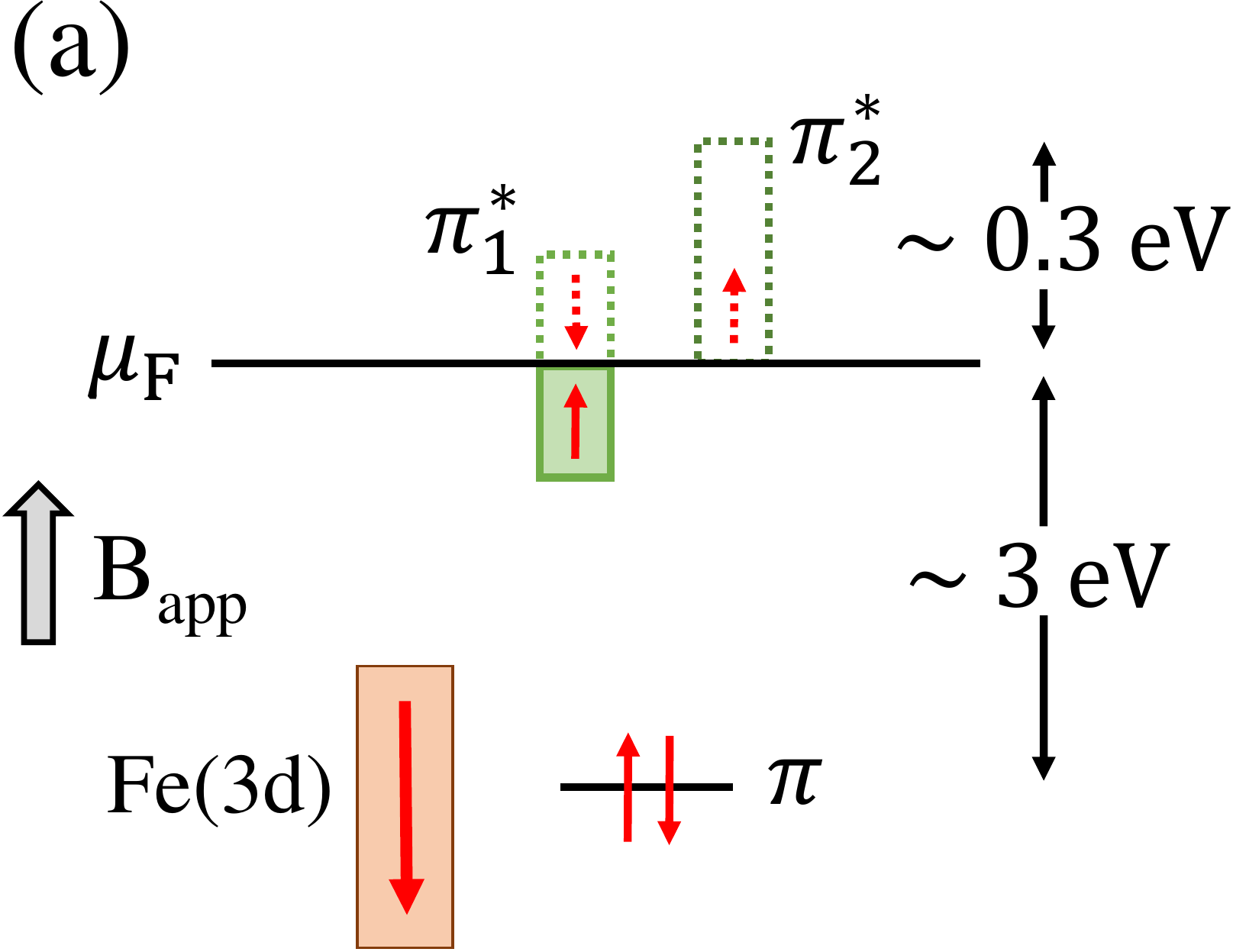}\vspace{1cm}
\justify
(a) Illustration of the spin polarizations for the Fe($3d$), 
and the bonding $\pi$ and the antibonding $\pi_1^*$ and $\pi_2^*$ host states
in an applied magnetic field ${\bf B}_{\rm app}$ for deoxy-heme.
For ${\bf B}_{\rm app}$ 
pointing up,
the total Fe($3d$) spin gets polarized in the down direction,
while the spin of the nearly half-filled $\pi_1^*$ state gets 
polarized parallel to the field. 
The $\pi_2^*$ state is nearly empty. 
From the DFT+QMC results we know that 
the $\pi$ states are located about 3 eV below the Fermi level,
and the widths of the $\pi_1^*$ and $\pi_2^*$ states are about 0.3 eV. 
\label{fig4a}
\end{figure}

\clearpage

\begin{figure}
\centering
\vspace{0.0cm}
\includegraphics[height=5.0cm]{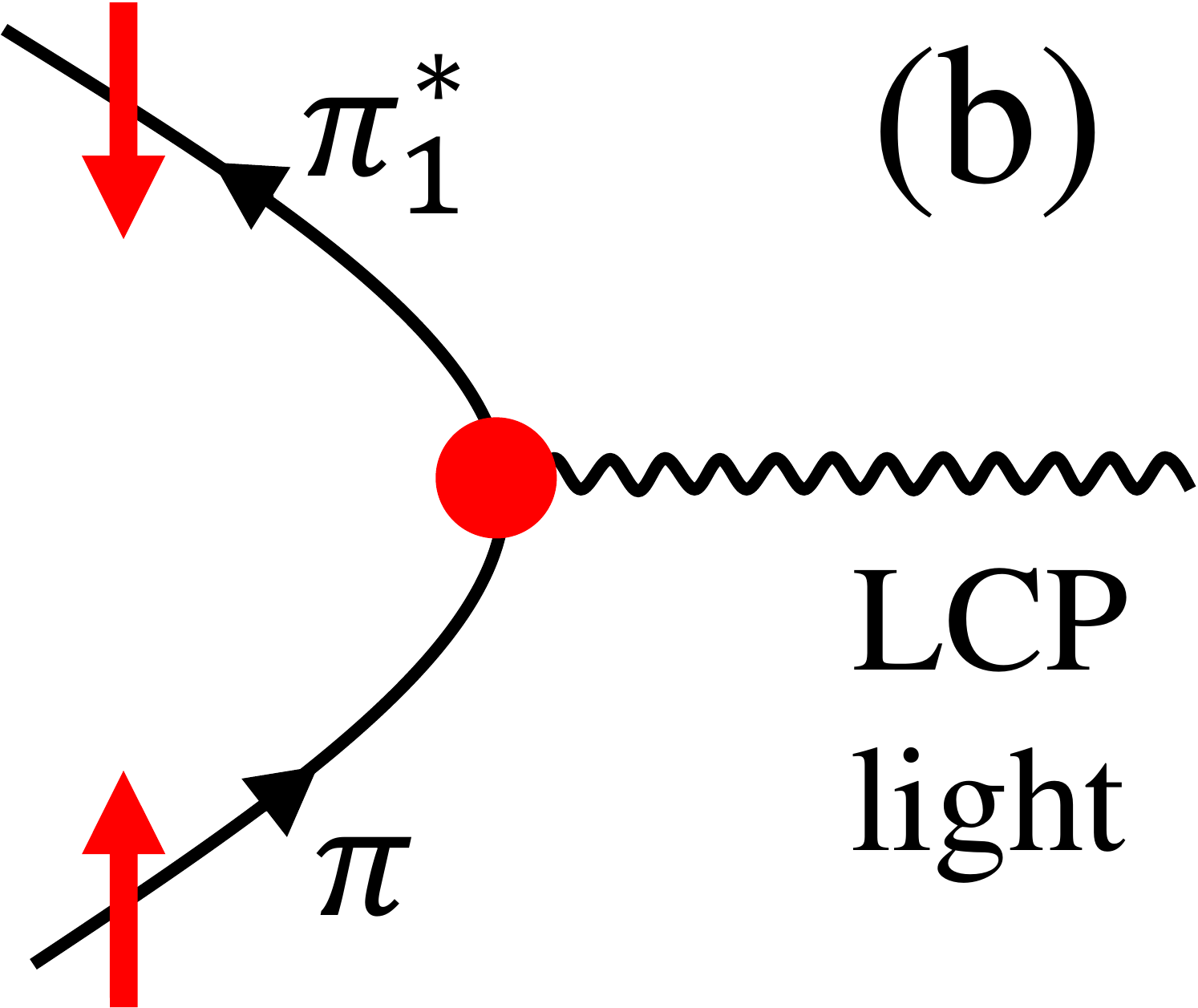}\hspace{2cm}
\includegraphics[height=5.0cm]{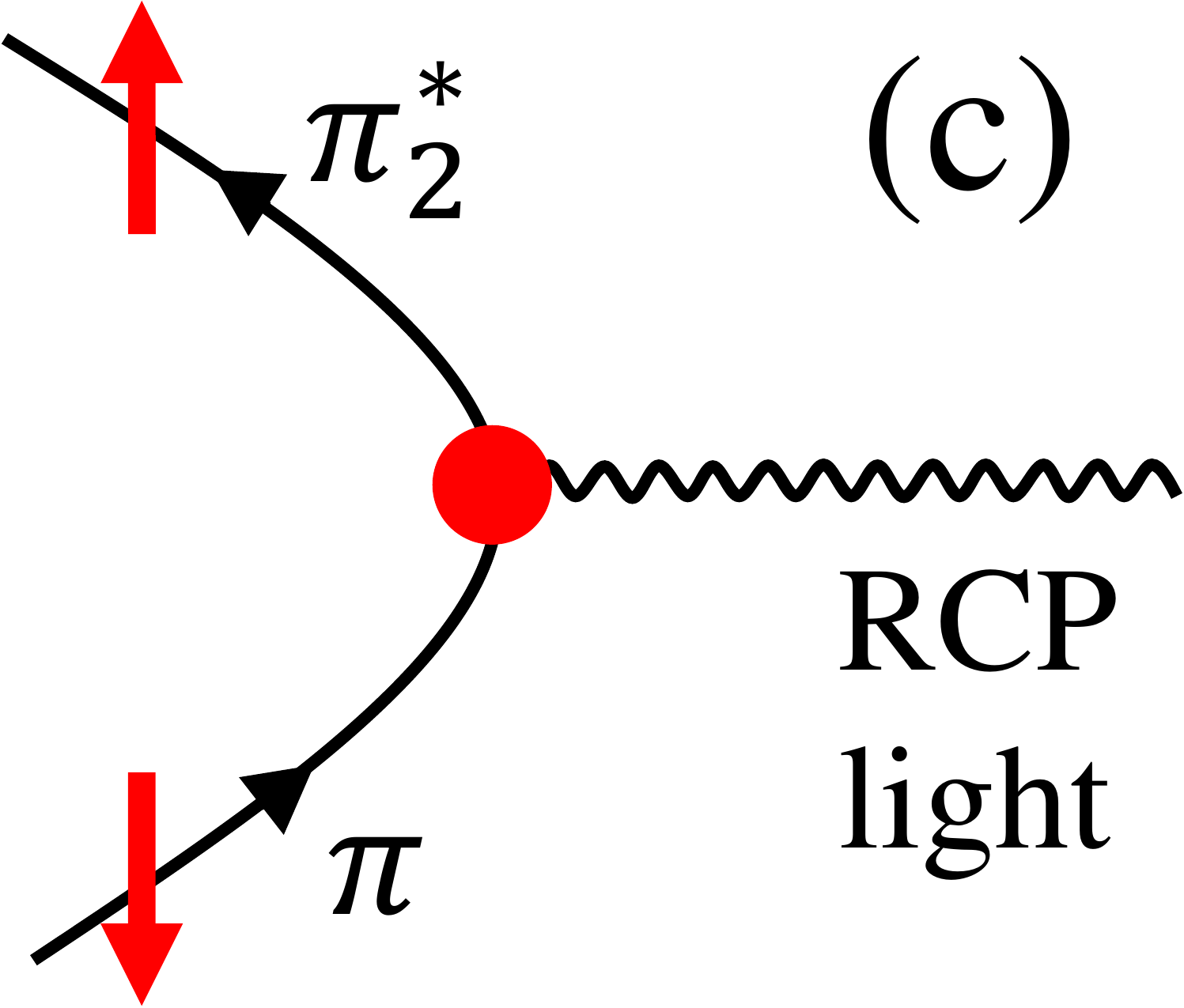}\vspace{1cm}
\justify
(b) Feynman diagram representing the absorption of LCP light
in the optical transition $\pi\rightarrow \pi_1^*$.
Here, an up-spin electron in a $\pi$ state makes a transition 
to the $\pi_1^*$ state becoming down-spin by 
absorbing LCP light (denoted by the wavy line).
In (c), the $\pi\rightarrow \pi_2^*$ transition is shown for the absorption 
of RCP light,
where a down-spin $\pi$ electron makes a transition 
to the $\pi_2^*$ state becoming up-spin. 
Here, the red dot is the effective vertex for the optical transition. 
Even though there is spin-orbit interaction only 
at the Fe($3d$) orbitals,
the $\pi$ states can gain an effective spin-orbit coupling through 
antiferromagnetic coupling and 
hybridization with the Fe($3d$) orbitals,
which are discussed in the Methods section.
\label{fig4bc}
\end{figure}

\clearpage

\begin{figure}
\centering
\vspace{0.0cm}
\includegraphics[width=13cm]{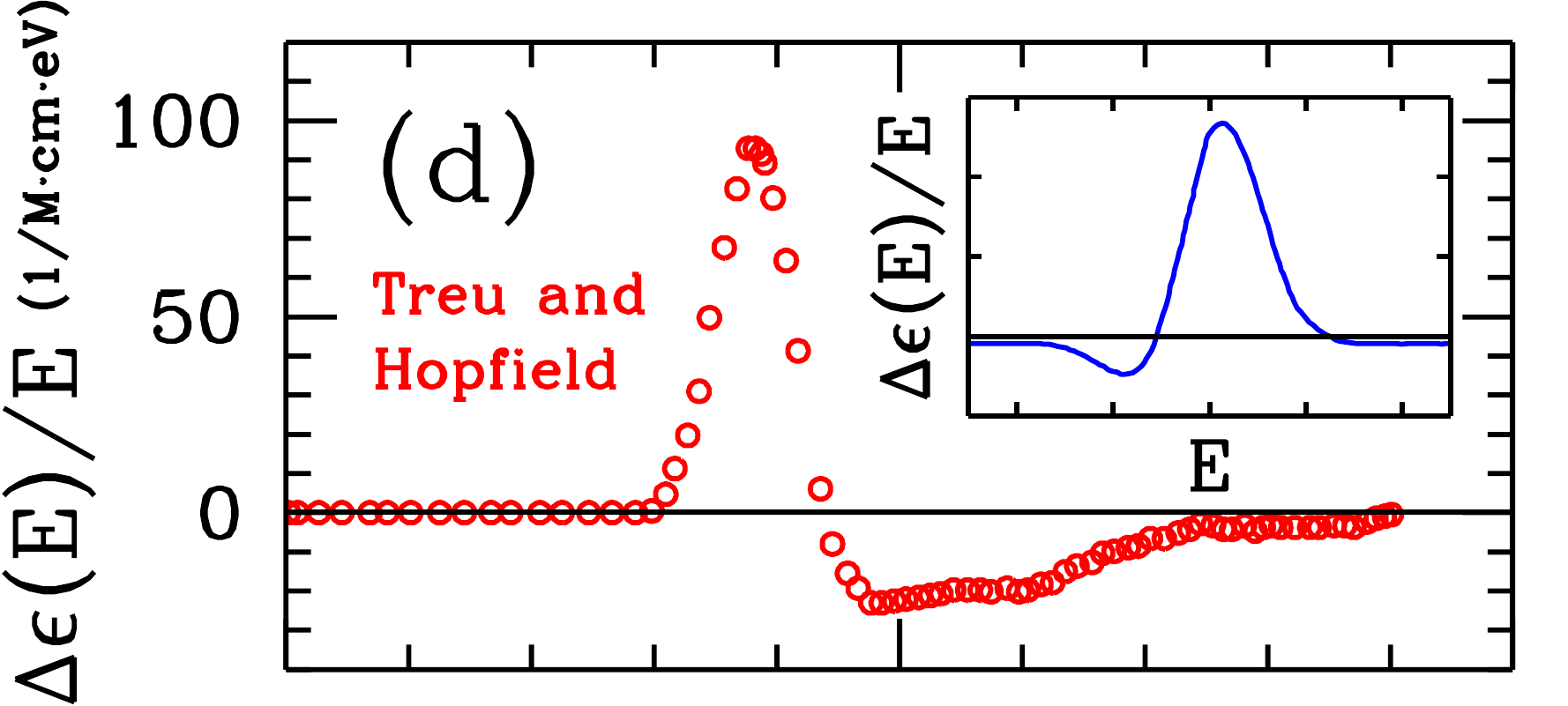}\vspace{0.5cm}
\includegraphics[width=13cm]{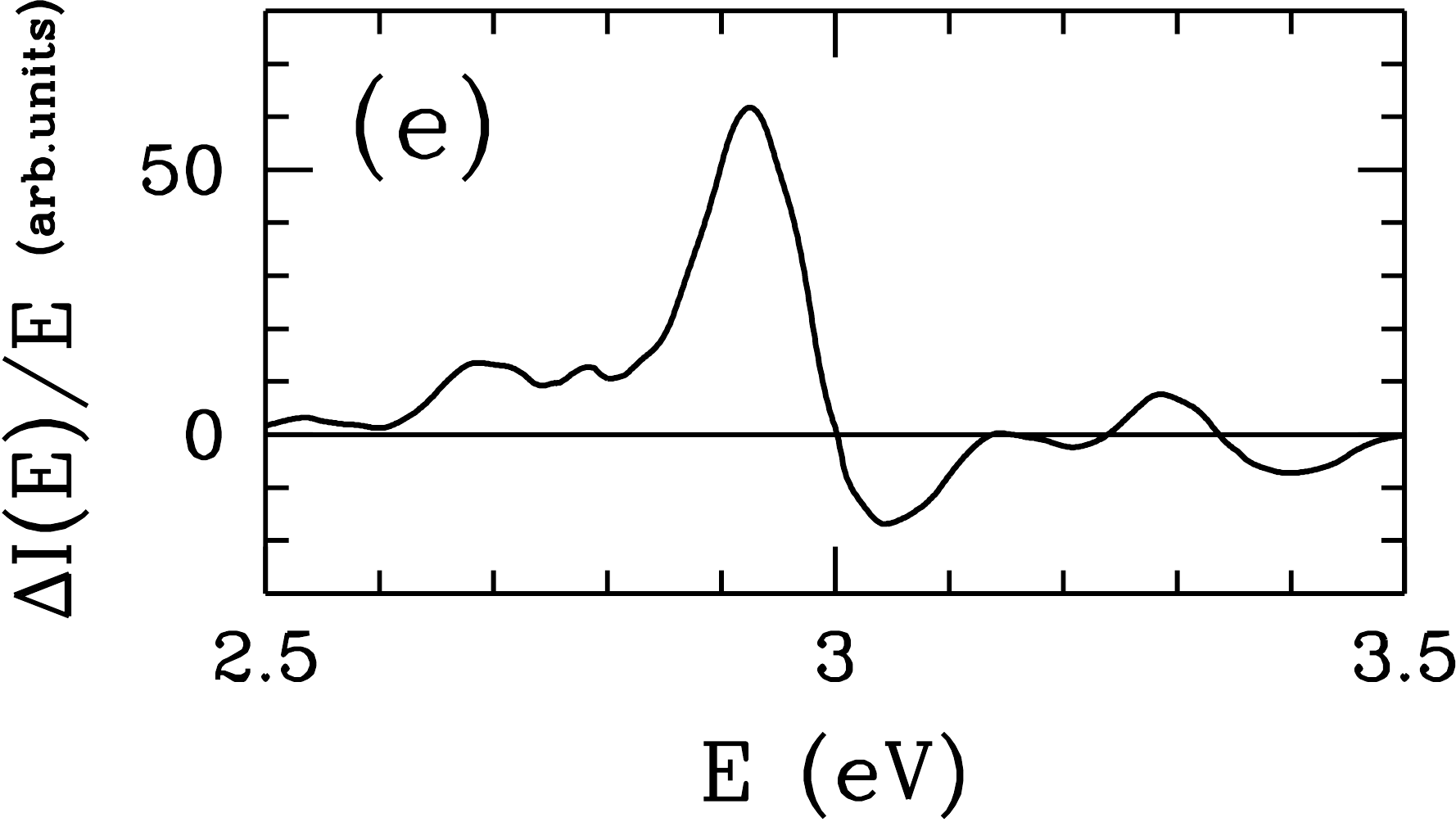}\vspace{1cm}
\justify
(d) Anomalous MCD line shape in the UV region for deoxy-HbA 
from the experiments by Treu and Hopfield \cite{Treu}. 
Inset: Line shape normally expected 
for the temperature-dependent MCD spectrum \cite{Mason}.
(e) Results from our calculation of the MCD spectrum for the deoxy-heme cluster.
\label{fig4de}
\end{figure}

\clearpage

%\begin{linenumbers}

{{\bf METHODS}}\\

{\bf Effective Anderson impurity model for heme} \\
We use an effective multi-orbital Anderson impurity model \cite{Anderson,Haldane},
where the five Fe($3d$) orbitals are taken 
as the impurity states and 
the remaining orbitals are treated as the host states,
to describe the electronic properties of the deoxy 
and oxy-heme clusters. 
The multi-orbital Anderson Hamiltonian 
with the intra and inter-orbital Coulomb interactions 
is given by 
\begin{eqnarray}
H = \sum_{m,\sigma} (\varepsilon_m-\mu) c^{\dagger}_{m\sigma} c_{m\sigma} 
& + &
\sum_{\nu,\sigma} (\varepsilon_{d\nu}-\mu) 
d^{\dagger}_{\nu\sigma} d_{\nu\sigma}  
+ \sum_{m,\nu,\sigma} ( V_{m\nu} c^{\dagger}_{m\sigma} d_{\nu\sigma} + 
V^*_{m\nu} d^{\dagger}_{\nu\sigma} c_{m\sigma} ) \\
& + &\sum_{\nu} U
n_{\nu\uparrow} n_{\nu\downarrow}
+ \sum_{\nu > \nu',\sigma} 
\big(U'n_{\nu\sigma} n_{\nu'-\sigma} + 
(U'-J) n_{\nu\sigma}n_{\nu'\sigma}\big), \nonumber 
\label{hamiltonian}
\end{eqnarray}
where $c^{\dagger}_{m\sigma}$ ($c_{m\sigma}$) operator creates (annihilates) an electron 
in the $m$'th host state with spin $\sigma$, 
$d^{\dagger}_{\nu \sigma}$ ($d_{\nu \sigma}$) is the creation
(annihilation) operator for a localized electron with spin $\sigma$ at 
the Fe($3d_{\nu}$) orbital, 
and the electron occupation operator for the Fe($3d_{\nu}$) orbitals is 
$n_{\nu\sigma}= d^{\dagger}_{\nu\sigma} d_{\nu\sigma}$.
The energies of the host and the Fe($3d_{\nu}$) states are
$\varepsilon_m$ and $\varepsilon_{d\nu}$,
respectively.
The hybridization matrix element 
between the $m$'th host state and the Fe($3d_{\nu}$) orbital is $V_{m\nu}$.
The intra-orbital Coulomb repulsion is $U$,
while 
$U'$ and $U''=U'-J$ are the Coulomb interactions between two $3d$
electrons in different orbitals with antiparallel and
parallel spins, respectively.
Here, $J$ is 
the ferromagnetic Hund's coupling constant.
In the case of a free atom,
the relation $U'=U-2J$ holds,
which we also use here.
The chemical potential $\mu$ is introduced  
because the QMC calculations are performed 
at finite temperatures in the grand canonical ensemble.
At each value of the temperature,
we adjust $\mu$ so that the cluster has the correct number of electrons,
which is discussed below in Extended Data Fig. 3. 
We obtain the values of 
$\varepsilon_m$, $\varepsilon_{d\nu}$ and $V_{m\nu}$
by the density-functional theory (DFT) \cite{Kohn},
which is discussed in the Supplementary Information. 
For the interaction parameters we use  $U=4$ eV
and $J=0.9$ eV. 

We have performed the DFT+QMC calculations for clusters obtained 
by truncating the full deoxy and oxy-heme molecular structures 
from the Protein Data Bank
as described in the Supplementary Information. 
For understanding the functioning of HbA,
the role of the stereochemical effects have been investigated. 
In particular,
it has been emphasized that Fe moves by about 0.4 {\AA} towards 
the porphyrin ring upon O$_2$ binding \cite{Perutz1960}.
Since we are using coordinates determined by the X-ray measurements,
these stereochemical effects are already included in our model. 

In Eq. (1), we include the longitudinal component of the 
Hund's interaction, 
however the transverse component, which consists of 
the spin-flip and the pair-hopping terms,
is not included. 
In addition, 
we neglect the temperature dependence of the molecular coordinates. 
In spite of these approximations,
the DFT+QMC technique applied to this effective
impurity model offers a realistic treatment for the 
electronic state of HbA. 
Comparisons with the experimental data on the magnetic properties show
that we capture the connection 
between the magnetism and the biological functioning 
at least at a qualitative level. 
We note that these DFT+QMC calculations 
represent the only computational approach which yields agreement 
with the magnetic susceptibility,
M\"ossbauer, and the MCD data on HbA at the same time. 
 
{\bf Additional DFT+QMC data on the spin and 
charge distributions in the heme clusters} \\
Extended Data Tables I(a) and (b) show the electron occupation number 
$\langle n_{\nu}\rangle$ 
and the effective magnetic moment 
$M_{\nu}^{\rm eff}$ of the Fe($3d_{\nu}$) orbitals 
(in units of Bohr magnetons $\mu_{\rm B}$) 
for the deoxy and oxy-heme clusters.
Extended Data Table I(c) shows the total Fe($3d$) moment $M_{3d}$, 
the total host moment $M_{\rm h}$, 
and the total moment $M_{\rm t}$ of the cluster. 
In addition, the Fe($3d$)-host magnetization correlation function 
$\langle M^z_{3d} M^z_{\rm h}\rangle$ 
is shown (in $\mu_{\rm B}^2$). 
The values of the Fe($3d$), the host and the total effective spins
denoted by $S_{3d}$, $S_{\rm h}$ and $S$, respectively, 
are also listed. 
The Monte Carlo statistical errors in 
$\langle n_{\nu}\rangle$ and $M_{\nu}^{\rm eff}$ 
are less than 2 percent. 
The QMC finite-$\Delta\tau$ effects on these 
are discussed below in Extended Data Fig. 4.
In Extended Data Tables I(d)-(f),
results on the magnetization correlation function 
$\langle M^z_{\nu} M^z_{\nu'}\rangle$
are shown (in $\mu_{\rm B}^2$)
for the deoxy- and oxy-heme clusters at various temperatures.
Here, the statistical errors are less than 0.04 $\mu_{\rm B}^2$. 
Extended Data Table I(g) compares the total intra and 
inter-orbital magnetic correlations among the 
Fe($3d$) orbitals for the deoxy and the oxy cases.
In these quantities, the statistical errors are less 
than 0.1 $\mu_{\rm B}^2$.

For the deoxy case, 
we see that $\langle n_{\nu}\rangle$ 
is close to unity for all five of the $3d$ orbitals. 
The total number of the Fe($3d$) electrons 
$\langle n_{3d}\rangle = 5.38$ at $T=300$ K, 
hence Fe is in between the ferric (+3) 
than the ferrous (+2) states. 
These orbitals are nearly fully polarized where
the corresponding 
$M_{\nu}^{\rm eff}$'s are slightly less than 1 $\mu_{\rm B}$.
The total effective moment of the $3d$ orbitals 
$M_{3d}$ is 4.56 $\mu_{\rm B}$,
while for the host $M_{\rm h}=1.06 \, \mu_{\rm B}$. 
The total moment $M_{\rm t}=4.08\,\mu_{\rm B}$ is less than $M_{3d}$,
because of the Fe($3d$)-host antiferromagnetic correlations.
We find the effective Fe($3d$) spin $S_{3d}$ to be 1.83,
while the total spin $S=1.60$.
As $T$ is lowered from 300 K to 150 K,
these values for the electron occupations,
effective moments and spins show almost no variation in the deoxy case.

In the oxy-heme case, 
the total Fe($3d$) electron number is 5.83 at 300 K. 
As $T$ decreases to 150 K, $\langle n_{3d}\rangle$ increases to 6.07.
This is due to charge transfer from mainly the O$_2$ 
to the Fe($3d$) orbitals. 
In the oxy case at 150 K, 
we have $\langle n_{xy}\rangle = 1.88$,
hence this orbital is nearly doubly-occupied.
The remaining orbitals are closer to half-filling. 

At 300 K for the deoxy-heme $M_{3d}= 4.56 \,\mu_{\rm B}$,
while it is 2.50 $\mu_{\rm B}$
in the oxy case.
Upon lowering $T$ to 150 K,
$M_{3d}$ of oxy-heme decreases to 1.65 $\mu_{\rm B}$.
This drop in $M_{3d}$ 
is mainly due to the loss of the ferromagnetic correlations 
among the Fe($3d$) orbitals.
This is seen by comparing the results on
$\langle M^z_{\nu} M^z_{\nu'} \rangle$ 
shown in Extended Data Tables I(d)-(f)
for the deoxy and oxy-heme clusters.
We note that the total effective Fe($3d$) moment is obtained from
\begin{eqnarray}
\langle (M^z_{3d})^2\rangle = 
\sum_{\nu} \langle (M^z_{\nu})^2\rangle + 
\sum_{\nu,\nu'\neq \nu} \langle M^z_{\nu} M^z_{\nu'} \rangle.
\end{eqnarray}
Extended Data Table I(g) compares the components of this expression 
for the deoxy and oxy cases, where we clearly see 
the loss of the inter-orbital ferromagnetic correlations
and its effect on $M_{3d}$.

In Extended Data Fig. 1(a), 
we compare the total spin susceptibility of the cluster $\chi_{\rm t}$ 
with the Fe($3d$) susceptibility $\chi_{\rm Fe}$.
For the deoxy case,
we see that $\chi_{\rm t}$ is reduced with respect to $\chi_{\rm Fe}$
due to the Fe-porphyrin antiferromagnetic coupling. 
Here, 
we also observe that $\chi_{\rm Fe}$ obeys a perfect Curie $T$-dependence. 
In the oxy case, 
there is little difference between $\chi_{\rm t}$ and $\chi_{\rm Fe}$.
Both increase as $1/T$ for $T$ down to 300 K.
As $T$ decreases below 300 K,
a magnetic gap opens at the Fermi level which causes 
$\chi_{\rm Fe}$ and $\chi_{\rm t}$ to vanish. 

In Extended Data Fig. 1(b), 
$\chi_{\rm t}$ of the oxy-heme cluster is shown 
as a function of $\mu$ for various $T$. 
Here, we clearly observe the development of a spin gap 
with $2\Delta_{\rm S} \approx 0.3$ eV at low $T$. 

{\bf Additional DFT+QMC data on the O$_2$ to Fe charge transfer 
and the magnetic gap in oxy-heme} \\
In Extended Data Fig. 2(a), 
we see that the total electron number in the Fe($3d$) orbitals
$\langle n_{3d}\rangle$ 
develops a peak at the Fermi level $\mu_{\rm F}\approx -3.9$ eV
at low $T$.
Extended Data Fig. 2(b) shows 
$\langle n_{\nu}\rangle$ versus $\mu$ 
for the five Fe($3d_{\nu}$) orbitals.
We observe that, 
if $\mu$ is kept fixed at $-3.9$ eV 
as $T$ is lowered from 400 K to 200 K, then 
$\langle n_{\nu}\rangle$ for the $t_{2g}$ orbitals 
($\nu=xy$, $xz$, and $yz$) gets enhanced.
In this case, 
the $3d_{xy}$ orbital has the largest increase. 
Meanwhile, for the $e_g$ orbitals ($\nu=3z^2-r^2$ and $x^2-y^2$) 
$\langle n_{\nu}\rangle$ gets suppressed. 
These show that the $3d$ orbitals are strongly coupled to each other
as well as to their environment. 
We also observe that 
the upper Hubbard level of the $3d_{xy}$ orbital 
shifts down in energy with decreasing $T$. 
We are able to capture these effects because we treat
Fe in heme as a Hund's impurity in the special electronic environment of oxy-heme 
by applying DFT+QMC.
We think that the system is minimizing its energy by redistributing the 
electrons and developing magnetic correlations. 
These lead us to the notion that the binding of O$_2$ to Fe in heme 
is related to magnetism and the $3d$ Coulomb interactions. 

In Extended Data Fig. 2(c),
we see that, 
as the peak in $\langle n_{3d}\rangle$ develops,
the total $3d$ magnetic  moment 
$M_{3d}$ gets suppressed at the Fermi level.
At low $T$,
$M_{3d}$ approaches 1.65 $\mu_{\rm B}$,
which is close to the spin-1/2 value. 

The electrons transferred to the Fe($3d$) orbitals
come mostly from a host state 
which we label as the $m=171$'th host state.
This state consists of the atomic orbitals
from O$_2$ and the proximal histidine.
Its wave function is illustrated in Extended Data Fig. 2(d). 
The electron occupation $\langle n_m\rangle$ of this state gets suppressed 
near the Fermi level at low temperatures,
while its magnetic moment $M_m$ gets enhanced reaching 0.6 $\mu_{\rm B}$ at 150 K
as shown in Extended Data Figs. 2(e) and (f), respectively. 
In Extended Data Table I(c),
we saw that 
the total host moment $M_{\rm h}\approx 1.5 \,\mu_{\rm B}$
at $T=150$ K. 
Hence, as seen in Fig. 2(b),
the total spin $S=0$ state develops because of the 
antiferromagnetic correlations between 
basically a spin-1/2 localized at the Fe site
and another spin-1/2 spread out at the neighboring 
O$_2$ and N sites. 

It is important to note that 
the $S=0$ spin state is possible because the electronic structure of 
oxy-heme is such that $\mu_{\rm F}$ is located close to the upper-Hubbard level of the 
Fe($3d_{xy}$) orbital. 
When the Fermi level is away from the upper-Hubbard level of the 
$3d_{xy}$ orbital,
we do not observe the charge transfer from O$_2$ to Fe and the magnetic gap. 
We also note that as the magnetic gap opens,
basically no changes occur in the electron occupation numbers and the magnetic moments 
at the nitrogen sites which are nearest-neighbor to Fe.
The magnetic moments at these nitrogen sites have already formed 
at higher temperatures. 

The O$_2$ to Fe charge transfer 
and the opening of the spin gap are 
interesting electronic properties of oxy-heme. 
This is actually reminiscent of the Kondo effect 
where a localized magnetic moment in a metallic host gets completely
screened by the formation of 
an antiferromagnetic screening cloud.
In that case, the metallic host develops a magnetic moment by bringing 
in an electron from the boundary of the system at infinity. 
However, 
in the oxy-heme cluster,
the host develops part of the screening moment by transferring electrons 
from the O$_2$ to the Fe($3d$) orbitals. 

{\bf Shift of the Fermi level with the temperature in oxy-heme} \\ 
In Extended Data Fig. 3, we show results on the total electron number 
$\langle n_{\rm t}\rangle$ versus $\mu$ for the oxy-heme cluster,
which has 77 atoms and 350 electrons.
We see that at 300 K 
the Fermi level $\mu_{\rm F}\approx -3.8$ eV, 
while at $T=150$ K we have $\mu_{\rm F}\approx -3.9$ eV.

{\bf QMC finite-$\Delta\tau$ effects in the oxy-heme case}\\  
The Matsubara-time step $\Delta\tau$ is introduced in the QMC method 
by dividing the inverse temperature $\beta=1/T$ into $L$ 
Matsubara-time slices, $\beta = L \, \Delta\tau$. 
The QMC results become exact in the limit $\Delta\tau\rightarrow 0$. 
All of the DFT+QMC results shown in the above figures were taken 
by using $\Delta\tau$ values between 0.125 and 0.2 eV$^{-1}$. 
In Extended Data Figures 4(a)-(d) we show the effects 
of using finite $\Delta\tau$
on $\langle n_{\rm t}\rangle$, $\langle n_{3d}\rangle$, 
$\chi_{\rm t}$ and $M_{\rm t}$, respectively.

The charge-neutral oxy-heme cluster has 350 electrons. 
Extended Data Fig. 4(a) shows that 
$\langle n_{\rm t}\rangle$ approaches 350.0
in the limit $\Delta\tau\rightarrow 0$ at $T=150$ K.
This result was obtained by using $\mu=-3.89$ eV. 
Hence, in the limit $\Delta\tau\rightarrow 0$ and at $T=150$ K
we have $\mu_{\rm F}\approx -3.9$ eV. 
For 200 K, $\mu=-3.89$ eV was also used,
while at $T=300$ K we used $\mu=-3.8$ eV.
These show that 
$\mu_{\rm F}$ shifts 
from $\approx -3.8$ eV at 300 K to $\approx -3.9$ eV at 150 K
with the opening of the spin gap. 

In Extended Data Fig. 4(b), we see that, 
in the limit $\Delta\tau\rightarrow 0$,
$\langle n_{3d}\rangle$ increases 
from 5.9 to approximately 6.1
as $T$ decreases from 300 K to 150 K.
Hence, Fe in oxy-heme is in a ferrous state.

Extended Data Fig. 4(c) shows that 
$\chi_{\rm t}\rightarrow 150\,\mu_{\rm B}^2$  
as $\Delta\tau\rightarrow 0$ at 300 K,
while for 200 K we see that $\chi_{\rm t}$ extrapolates 
to a much reduced value.
Hence,
the magnetic gap is already developed at 200 K.
When the $\Delta\tau\rightarrow 0$ corrections are taken into account,
the drop in $\chi_{\rm t}$ for $T< 300\,{\rm K}$,
which was seen in Fig. 3(a), becomes more rapid.
Similarly, in Extended Data Fig. 4(d) we observe 
a rapid drop in $M_{\rm t}$
in the limit $\Delta\tau \rightarrow 0$
as $T$ decreases. 
These figures show that the finite-$\Delta\tau$ 
effects do not change our conclusions. 

{\bf Calculation of the MCD spectrum in the UV region for deoxy-heme} \\ 
Extended Data Fig. 5(a) illustrates the initial, intermediate and the final states 
in the $\pi \rightarrow \pi_1^*$ optical transition with LCP light absorption
when the applied field is in the up direction. 
The left panel shows the initial state 
where the Fe($3d_{\nu}$) orbital, 
which is hybridizing with the bonding $\pi$ state, 
has a down-spin electron, 
while the $\pi$ state is doubly occupied. 
Because of the Fe($3d$)-$\pi^*$ antiferromagnetic correlations,
the antibonding $\pi_1^*$ state has an up-spin electron.
In the intermediate state (middle panel),
we see that,
upon LCP light absorption, the down-spin $\pi$ electron
moves to the $\pi_1^*$ state,
while the up-spin $\pi$ electron 
spin-flips together with the Fe($3d$) down-spin electron
through antiferromagnetic exchange. 
In this intermediate state, 
the up-spin Fe($3d$) electron is in an excited state
because of the ferromagnetic Hund's coupling to the other Fe($3d$) spins. 
Hence, 
it spin-flips one more time 
by making use of the spin-orbit coupling at the Fe site,
which leads to the final state (right panel). 
There is a similar process for the $\pi \rightarrow \pi_2^*$ transition
with the RCP light absorption.
These optical transitions are orbital selective
because of the antiferromagnetic Fe-$\pi^*$ correlations,
which leads to the anomalous MCD line shape.

From comparisons with the optical absorption data, 
which will be shown below in Extended Data Fig. 7,
we deduce that a bonding $\pi$ state located at -5.6 eV,
which we label as $\pi_1$,
is dominant in the UV optical processes. 
In fact,
we find that the leading contribution to UV MCD arises from 
the $\pi_1\rightarrow \pi_1^*$ transition for the LCP light absorption,
and $\pi_1\rightarrow \pi_2^*$ for RCP absorption. 
Extended Data Figs. 5(b)-(d) show 
the wave functions of these $\pi_1$, $\pi_1^*$, and  $\pi_2^*$ states,
which consist of the C($2p_z$) atomic orbitals of the porphyrin layer. 

Even though the $\pi$ states do not have significant spin-orbit coupling, 
they can gain an effective coupling 
because of antiferromagnetic correlations and hybridization 
with the Fe($3d$) orbitals.
Extended Data Figs. 5(e)-(f) illustrate the lowest-order process
where a $\pi$ state can gain an effective spin-orbit coupling.
We note that the largest hybridization matrix elements of the 
$\pi_1^*$ and $\pi_2^*$ states are with the $3d_{xz}$ and $3d_{yz}$ orbitals,
respectively, and they are both about 0.3 eV. 

We use the following simple expression
for the $T$-dependent MCD spectrum in the UV region
\begin{eqnarray}
\Delta I(E) = \sum_{m} 
\bigg(&C_{1}&  \,
|\langle \pi_1^* |{\bf p}| m \rangle|^{2}\,
A_{\pi^{*}_{1}}(\epsilon_m +E) \\
&-&  C_{2} \,
|\langle \pi_2^* |{\bf p}| m \rangle|^{2} \,
A_{\pi^{*}_{2}}(\epsilon_m +E) \bigg) 
f(\epsilon_m) (1-f(\epsilon_m +E)), \nonumber  
\end{eqnarray}
which is due to LCP $\pi\rightarrow \pi_1^*$ and 
RCP $\pi\rightarrow \pi_2^*$ optical transitions.
Here, $m$ sums over all of the bonding $\pi$ host states,
and $C_1$ and $C_2$ are coefficients to be determined.
In addition, we use
the matrix element
$|\langle m'|{\bf p}|m\rangle|^{2}$,
instead of 
$|\langle m'|p_-|m\rangle|^{2}$ or 
$|\langle m'|p_+|m\rangle|^{2}$
where $p_{\pm}=p_x\pm ip_y$,
since the experiments were performed
using samples in solution.

In Extended Data Fig. 6(a)-(d),
we show QMC data on the electronic properties of 
the $\pi_1^*$ and $\pi_2^*$ host states.
Their spectral weights $A(\omega)$ 
are obtained in an approximate way by taking the derivative 
with respect to $\mu$
of their electron occupations $\langle n\rangle$
shown in Extended Data Fig. 6(a).
The resulting
$A_{\pi^*_1}(\omega)$ and $A_{\pi^*_2}(\omega)$
are shown in Extended Data Fig. 6(b). 
These spectral weights have peaks split by 0.2 eV 
corresponding to $2J_{\rm AF}$,
where $J_{\rm AF}\sim 4|V_{\pi^*,\nu}|^2/U\sim 0.1\,eV$
is the maximum value of the 
Fe($3d_{\nu}$)-$\pi^*$
antiferromagnetic exchange constant. 
Extended Data Fig. 6(c) shows 
$\langle (M^z)^2\rangle$ versus $\mu$ for $\pi_1^*$ and $\pi_2^*$.
Extended Data Fig. 6(d) shows the function
$\langle M_{3d}^z M_m^z \rangle $
between Fe($3d$) and the $m=\pi_1^*$ and $\pi_2^*$ states. 
Here, we clearly see the Fe($3d$)-$\pi^*$ antiferromagnetic correlations.

In Eq. (3)
we take the ratio $C_1/C_2$ to be 4. 
This is the only fitting parameter we use
in obtaining the spectrum shown in Fig. 4(e). 
Since $\pi_1^*$ is already occupied by one up-spin electron 
and $\pi_2^*$ is nearly empty,
we expect the coefficient $C_1$ to be larger than $C_2$.
Finally, since the polarizability of the total Fe($3d$) spin 
has $1/T$ temperature dependence,
our result for the MCD intensity 
also exhibits this type $T$-dependence. 

{\bf Comparison of the MCD and the optical absorption data 
in the UV region for deoxy heme} \\
Extended Data Fig. 7(a) shows experimental data
on the optical absorption of deoxy-heme \cite{Treu}.
The peak near 3 eV 
originates from $\pi\rightarrow \pi^*$ transitions. 
This peak has an asymmetrical line shape,
where the intensity below 2.8 eV is much reduced
compared to the high energy tail. 

A simple estimate of the optical absorption in the UV region is given by
\begin{eqnarray}
I(E) = \sum_{m',m} 
|\langle m'|{\bf p}|m\rangle|^{2}
f(\epsilon_{m}) (1-f(\epsilon_{m'})) 
\delta(E - (\epsilon_{m'}-\epsilon_{m})),
\label{IE}
\end{eqnarray}
where $m$ and $m'$ sum over the bonding $\pi$ and 
antibonding $\pi^*$ host states, respectively,
and ${\bf p}=\sum_i\, (-e){\bf r}_i$ 
is the operator for the electric-dipole moment of the cluster. 
Here, for $\varepsilon_m$
we have used the DFT results, 
since the Coulomb interactions taken into account by QMC 
do not produce a significant shift in 
the doubly-occupied $\pi$ states.
The results for $I(E)/E$
are shown as the black bars for each transition
in Extended Data Fig. 7(b).
The solid curve was obtained
by using a broadening of $0.1$ eV for the $\delta$-functions.
Comparison with the experimental data 
shows that the dominant contribution to optical absorption
is coming from the 
$\pi_1\rightarrow \pi_1^*$ and $\pi_1\rightarrow \pi_2^*$ transitions,
where $\pi_1$ is located at -5.6 eV. 
We calculated the MCD spectra allowing only for these two transitions.
The result is the black curve shown in Extended Data Fig. 8,
which is in better agreement with the experimental data of Fig. 4(d). 

Extended Data Fig. 6(b) shows that 
the spectral-weight functions of 
the $\pi_1^*$ and $\pi_2^*$ states overlap in energy 
by about 0.2 eV.
We artificially eliminated this overlap 
by shifting the spectral weight of the $\pi_2^*$ state by 0.2 eV higher in energy. 
Its result
is shown as the red dashed curve in Extended Data Fig. 8.
Hence, we obtain 
the best agreement with the experimental MCD data 
when $\pi_1^*$ and $\pi_2^*$ spectral weights have no overlap. 

\vspace{0.8cm}
%{\bf Data availability}

%The data that support the findings of this study are available 
%from the corresponding author upon reasonable request.

%\end{linenumbers}
\clearpage

\begin{table}
\caption*{
{\bf Extended Data Table I $\vert$ Additional DFT+QMC data on the charge 
and spin distributions, and the magnetic correlations}
}
\vspace{0.5cm} 
\begin{tabular}{|c|C{1.5cm}C{1.5cm}C{1.5cm}C{1.5cm}C{1.5cm}C{2cm}|}
\hline
$\langle n_{\nu} \rangle$ & $xy$ & $xz$ & $yz$ & $3z^2-r^2$ & $x^2-y^2$
& total ($\langle n_d \rangle$) \\   
\hline 
 deoxy (300 K)           &   1.00  & 1.12  & 1.06  & 1.07  & 1.13  & 5.38 \\ 
\hline 
  oxy (300 K)            &   1.84  & 1.17  & 1.15  & 0.92  & 0.75  & 5.83 \\  
  oxy (150 K)            &   1.88  & 1.32  & 1.34  & 0.85  & 0.68  & 6.07 \\   
\hline
\end{tabular}
\vspace{0.2cm}
\justify
(a) Electron occupation number $\langle n_{\nu} \rangle$ of the Fe($3d_{\nu}$) orbitals.
\end{table}

\begin{table}
\vspace{0.5cm}
\begin{tabular}{|c|C{1.5cm}C{1.5cm}C{1.5cm}C{1.5cm}C{1.5cm}|}
\hline
$M^{\rm eff}_{\nu}\,\, (\mu_{\rm B})$  & $xy$ & $xz$ & $yz$ & $3z^2-r^2$ & $x^2-y^2$\\
\hline 
 deoxy (300 K)            & 0.99  & 0.93  & 0.96  & 0.95  & 0.93 \\ 
\hline 
  oxy (300 K)             & 0.40  & 0.89  & 0.89  & 0.79  & 0.74 \\
  oxy (150 K)             & 0.35  & 0.77  & 0.76  & 0.77  & 0.72 \\
\hline
\end{tabular}
\vspace{0.2cm}
\justify
(b) Effective magnetic moment $M^{\rm eff}_{\nu}$ of the Fe($3d_{\nu}$) orbitals.
\end{table}

\begin{table}
\vspace{0.5cm}
\begin{tabular}{|c|C{1.5cm}C{1.5cm}C{1.5cm}C{1.5cm}C{1.5cm}C{1.5cm}C{1.5cm}|}
\hline
 & $M_{3d}$  & $M_{\rm h}$ & $\langle M^z_{3d} M^z_{\rm h}\rangle$ &  $M_{\rm t}$ 
 & $S_{3d}$ & $S_{\rm h}$ 
 & $S$ \\
\hline
deoxy (300 K)           &  4.56   & 1.06 & -2.65  &  4.07  &  1.83   & 0.23 & 1.60 \\
\hline 
oxy (300 K)             &  2.50   & 1.32 & -1.71  &  2.13  &  0.85   & 0.33 & 0.68 \\
oxy (150 K)             &  1.65   & 1.50 & -2.30  &  0.61  &  0.47   & 0.40 & 0.10 \\
\hline
\end{tabular}
\vspace{0.2cm}
\justify
(c) Effective magnetic moments and effective spins. 
Here, $M_{3d}$, $M_{\rm h}$ and $M_{\rm t}$ denote the Fe($3d$), host and the total
effective magnetic moments (in units of $\mu_{\rm B}$), respectively,
while 
$\langle M^z_{3d} M^z_{\rm h}\rangle$ is the Fe($3d$)-host magnetic correlation function 
(in $\mu_{\rm B}^2$).
In addition, 
$S_{3d}$, $S_{\rm h}$ and $S$ denote the Fe($3d$), 
host and the total effective spins, 
respectively.
\end{table}

\clearpage

\begin{table}
\vspace{0.0cm}
\begin{tabular}{|c|C{1.5cm}C{1.5cm}C{1.5cm}C{1.5cm}C{1.5cm}|}
\hline
 $\langle M^z_{\nu} M^z_{\nu'} \rangle$ 
 & $xy$ & $xz$ & $yz$ & $3z^2-r^2$ & $x^2-y^2$ \\ 
\hline
$xy$         &   0.98   & 0.84  &  0.90  & 0.88   &  0.84  \\
$xz$         &          & 0.86  &  0.79  & 0.77   &  0.73  \\
$yz$         &          &       &  0.92  & 0.83   &  0.78  \\
$3z^2-r^2$   &          &       &        & 0.90   &  0.77  \\
$x^2-y^2$    &          &       &        &        &  0.86  \\
\hline
\end{tabular}
\vspace{0.2cm}
\justify
(d) Magnetic correlation function $\langle M^z_{\nu} M^z_{\nu'} \rangle$ 
among the Fe($3d_{\nu}$) orbitals for the deoxy-heme cluster at $T=300\,{\rm K}$.
\end{table}

\begin{table}
\vspace{0.5cm}
\begin{tabular}{|c|C{1.5cm}C{1.5cm}C{1.5cm}C{1.5cm}C{1.5cm}|}
\hline
 $\langle M^z_{\nu} M^z_{\nu'} \rangle$ 
 & $xy$ & $xz$ & $yz$ & $3z^2-r^2$ & $x^2-y^2$  \\ 
\hline
$xy$         &   0.16   & 0.06  &  0.06  & 0.04   &  0.01  \\
$xz$         &          & 0.79  &  0.60  & 0.25   &  0.14  \\
$yz$         &          &       &  0.81  & 0.25   &  0.14  \\
$3z^2-r^2$   &          &       &        & 0.62   &  0.08  \\
$x^2-y^2$    &          &       &        &        &  0.55  \\
\hline
\end{tabular}
\vspace{0.2cm}
\justify
(e) Magnetic correlation function $\langle M^z_{\nu} M^z_{\nu'} \rangle$ 
among the Fe($3d_{\nu}$) orbitals for the oxy-heme cluster at $T=300\,{\rm K}$.
\end{table}
 
\begin{table}
\vspace{0.5cm}
\begin{tabular}{|c|C{1.5cm}C{1.5cm}C{1.5cm}C{1.5cm}C{1.5cm}|}
\hline
 $\langle M^z_{\nu} M^z_{\nu'} \rangle$ 
 & $xy$ & $xz$ & $yz$ & $3z^2-r^2$ & $x^2-y^2$ \\ 
\hline
$xy$         &   0.12   & 0.02  &  0.02  & 0.02   &  0.00  \\
$xz$         &          & 0.59  & -0.09  & 0.05   &  0.04  \\
$yz$         &          &       &  0.58  & 0.04   &  0.04  \\
$3z^2-r^2$   &          &       &        & 0.59   &  0.04  \\
$x^2-y^2$    &          &       &        &        &  0.52  \\
\hline
\end{tabular}
\vspace{0.2cm}
\justify
(f) Magnetic correlation function $\langle M^z_{\nu} M^z_{\nu'} \rangle$ 
among the Fe($3d_{\nu}$) orbitals for the oxy-heme cluster at $T=150\,{\rm K}$.
\end{table}

\clearpage

\begin{table}
\vspace{0.0cm}
\begin{tabular}{|c|C{2cm}C{3cm}C{2cm}|}
\hline
& $\sum_{\nu} \langle(M^z_{\nu})^{2}\rangle$ & 
$\sum_{\nu \neq \nu'} \langle M^z_{\nu} M^z_{\nu'} \rangle$ & 
$\langle (M^z_{3d})^{2} \rangle$ \\
\hline
deoxy (300 K)            & 4.5        & 16.2   & 20.7 \\
\hline 
oxy (300 K)              & 2.9        & 3.3    & 6.2  \\
oxy (150 K)              & 2.4        & 0.3    & 2.7  \\
\hline
\end{tabular}
\vspace{0.2cm}
\justify
(g) Comparison of the intra and inter-orbital 
Fe($3d$) magnetic correlations (in units of $\mu_{\rm B}^2$)
contributing to $(M_{3d})^2$,
the square of the effective total Fe($3d$) moment. 
\end{table}

\clearpage

\begin{figure}
\centering
\caption*{
{\bf Extended Data Fig. 1 $\vert$ Additional DFT+QMC results on the
spin susceptibility}
\vspace{1.0cm}
}
{\includegraphics[width=7.5cm]{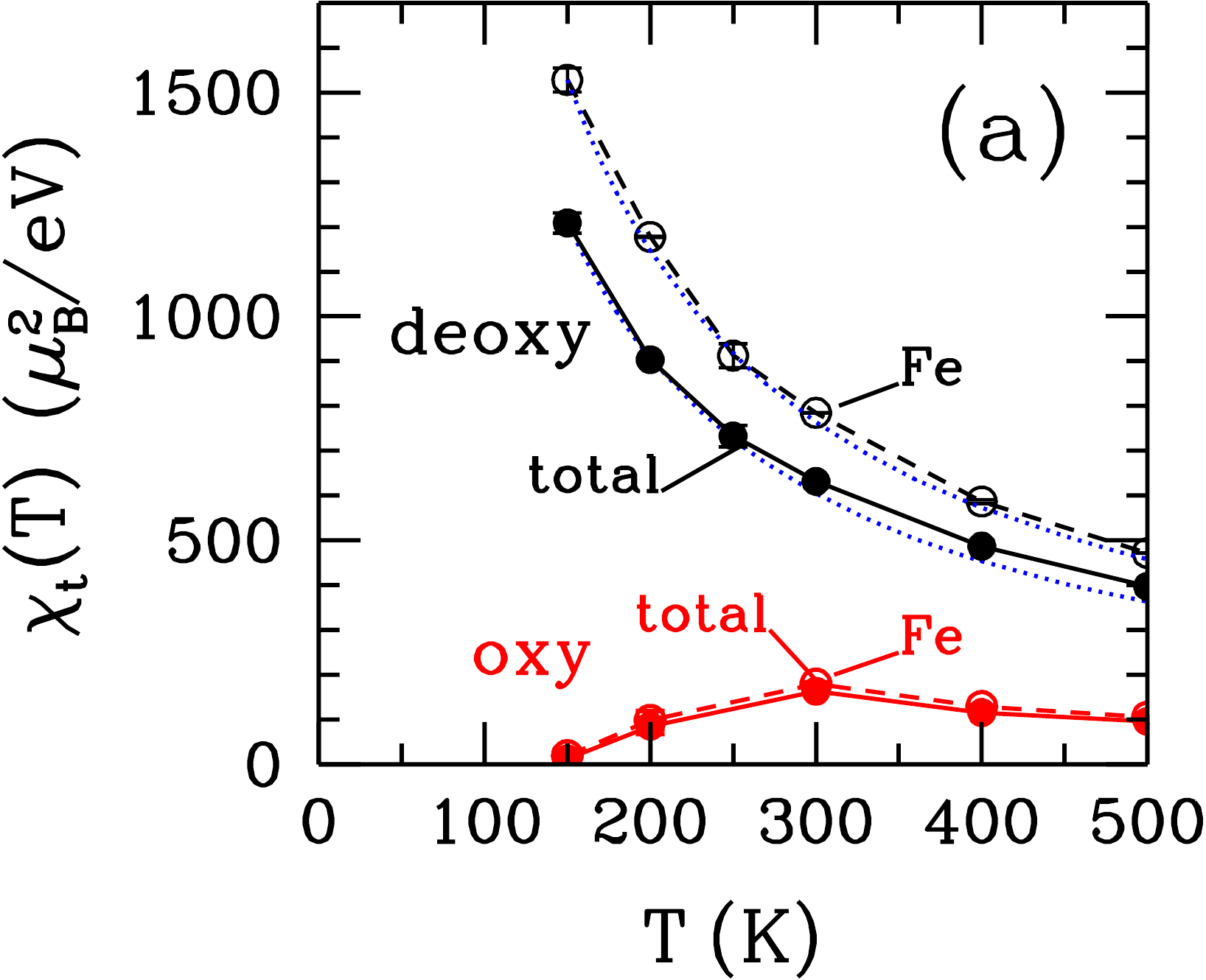}\hspace{1.0cm}}
{\includegraphics[width=7.5cm]{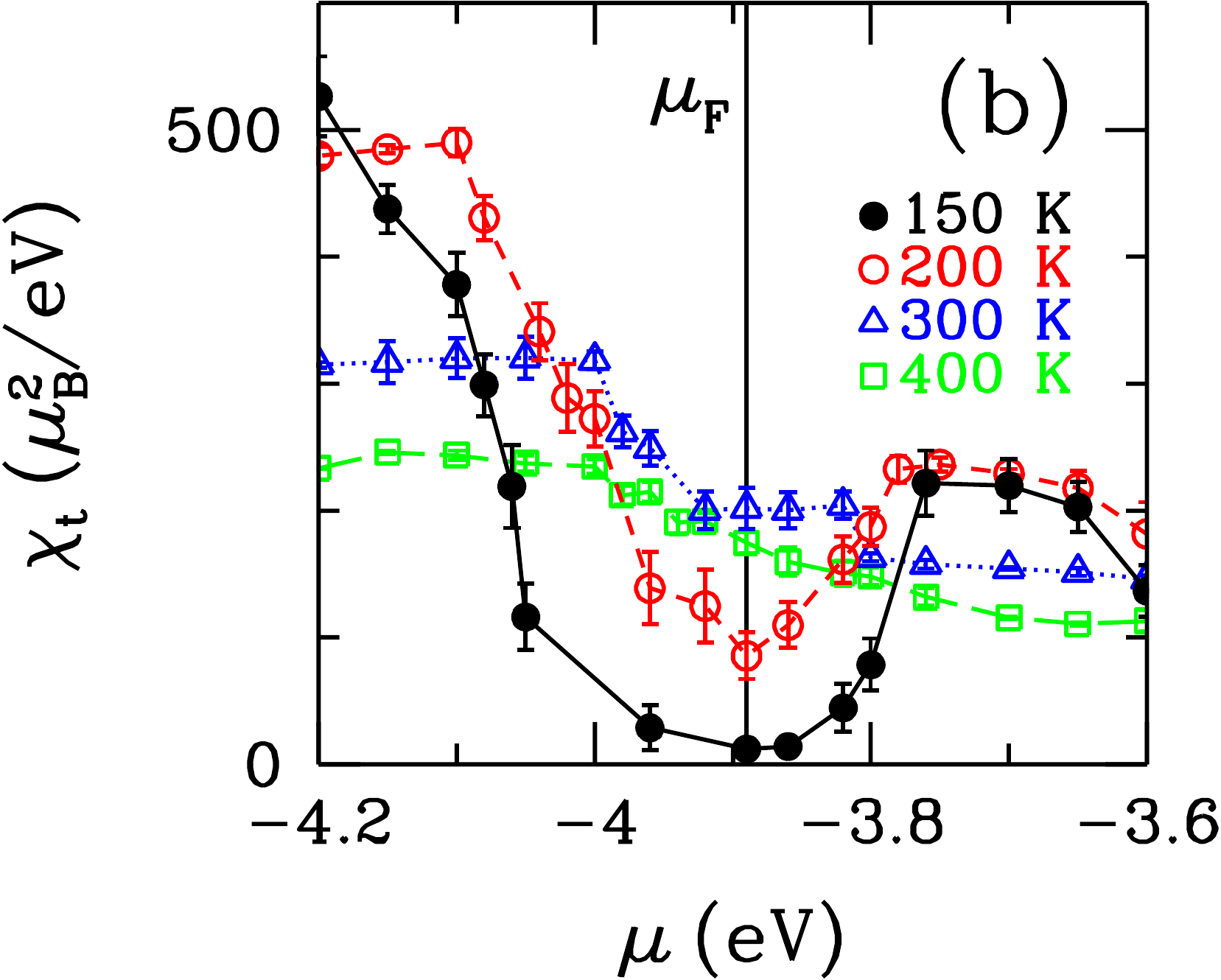}\vspace{1.0cm}}
\justify
(a) Total spin susceptibility $\chi_{\rm t}$ of the deoxy 
and oxy-heme clusters denoted by the filled circles plotted 
as a function of the temperature.
The results on the Fe spin susceptibility $\chi_{\rm Fe}$ 
denoted by the open circles
are also shown.
Here, the blue dotted lines denote the $1/T$ temperature dependence. 
(b) $\chi_{\rm t}$ versus the chemical potential $\mu$
near the Fermi level $\mu_{\rm F}$ at various temperatures. 
Here, the black vertical line denotes $\mu_{\rm F}$
at $T=150$ K. 
\label{XDfig1a}
\end{figure}
\clearpage

\begin{figure}
\centering
\caption*{
{\bf Extended Data Fig. 2 $\vert$ Additional DFT+QMC data 
on the charge transfer from O$_2$ to Fe in oxy-heme} 
}
\vspace{1cm}
{\includegraphics[width=7.5cm]{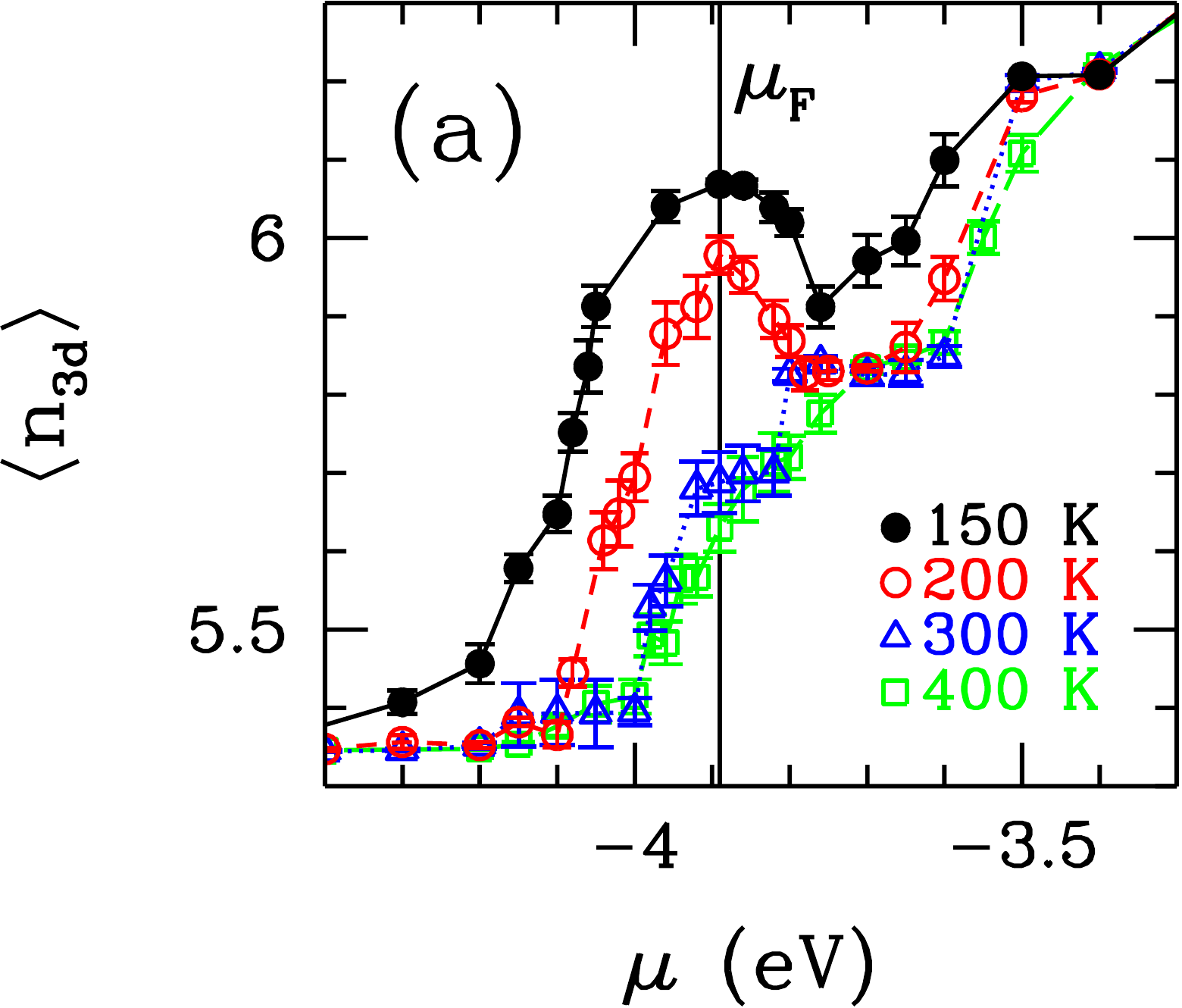}\hspace{0.5cm}}
{\includegraphics[width=7.5cm]{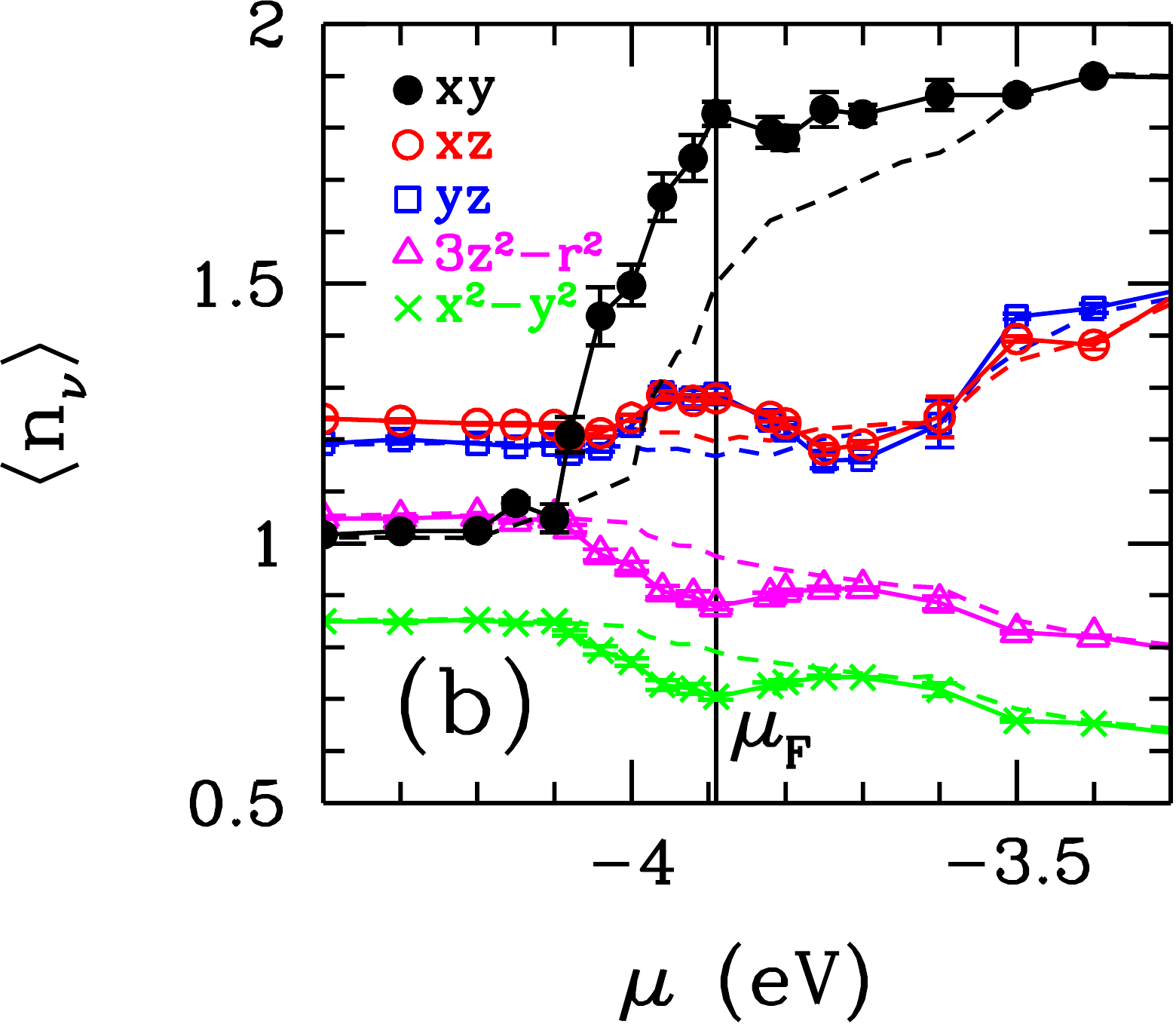}\vspace{1.0cm}}
{\includegraphics[width=7.5cm]{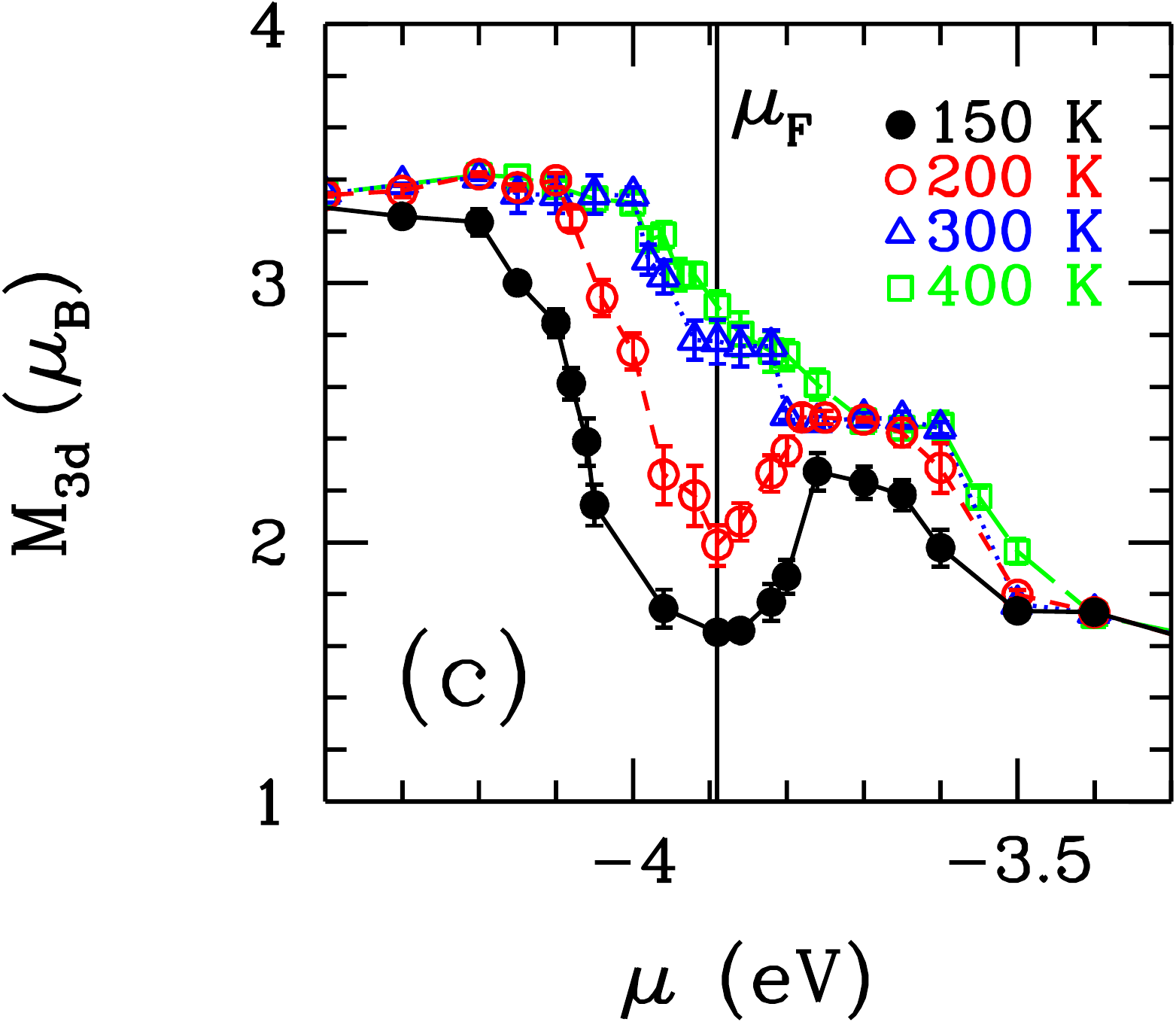}\hspace{3.0cm}}
{\includegraphics[width=4cm]{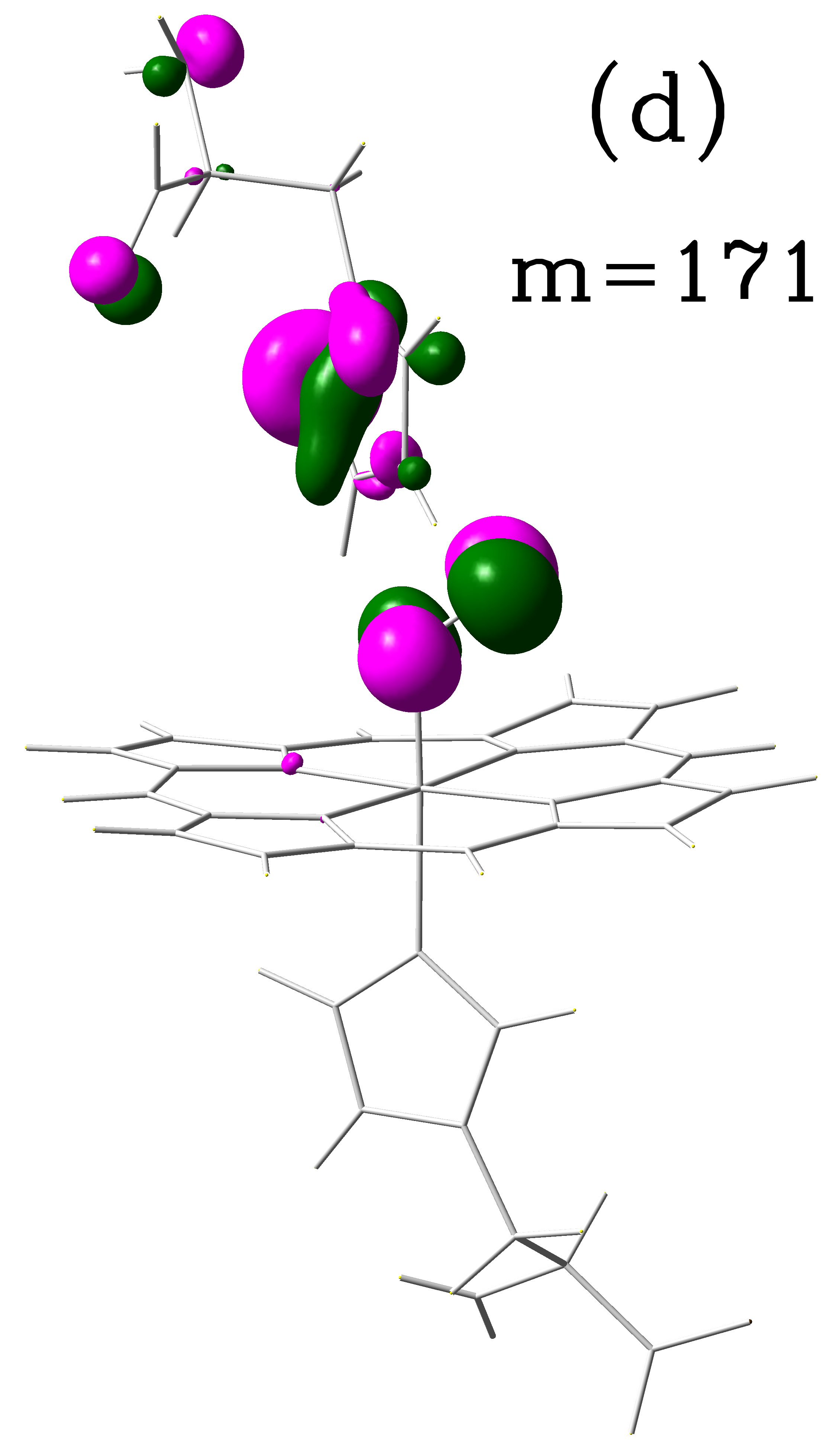}\hspace{0.9cm}\vspace{1.0cm}}
\justify
(a) Total number of the electrons in the Fe($3d$) orbitals 
$\langle n_{3d}\rangle$ plotted as a function of the chemical potential 
$\mu$ at various temperatures.
(b) Electron occupation number of the Fe($3d_{\nu}$) orbitals 
$\langle n_{\nu}\rangle$ versus $\mu$.
Here, the data points connected by the solid lines denote
$\langle n_{\nu}\rangle$ obtained at 200 K,
while the dashed curves denote results obtained at 400 K.
(c) Total magnetic moment of the Fe($3d$) orbitals 
$M_{3d}$ versus $\mu$.
(d) Illustration of the wave function of the $m=171$'th host state,
which is real valued.
Here, the magenta and green colors denote the positive and the negative regions. 
\label{XDfig2}
\end{figure}

\begin{figure}
\centering
{\includegraphics[width=7.5cm]{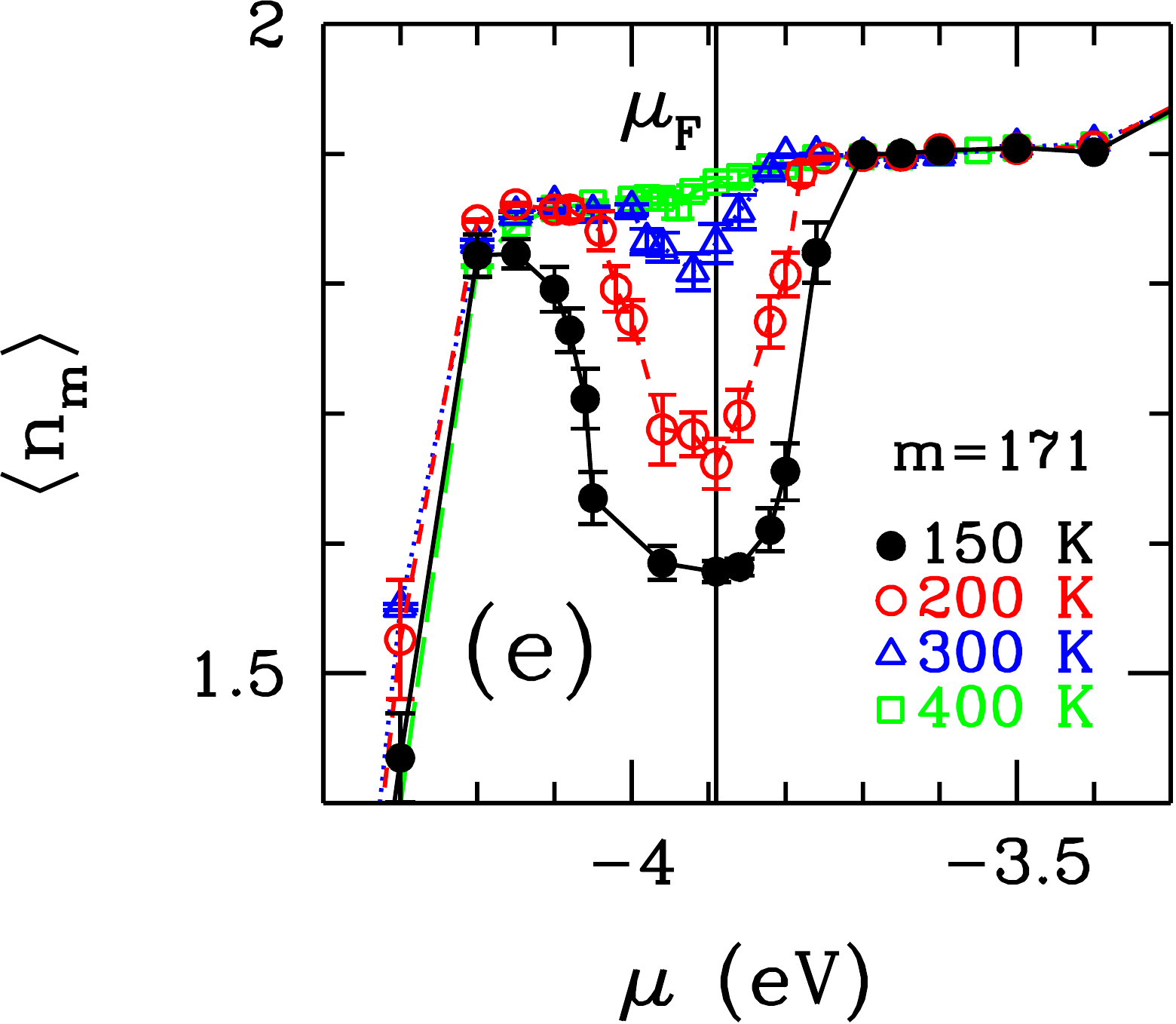}\hspace{0.5cm}}
{\includegraphics[width=7.5cm]{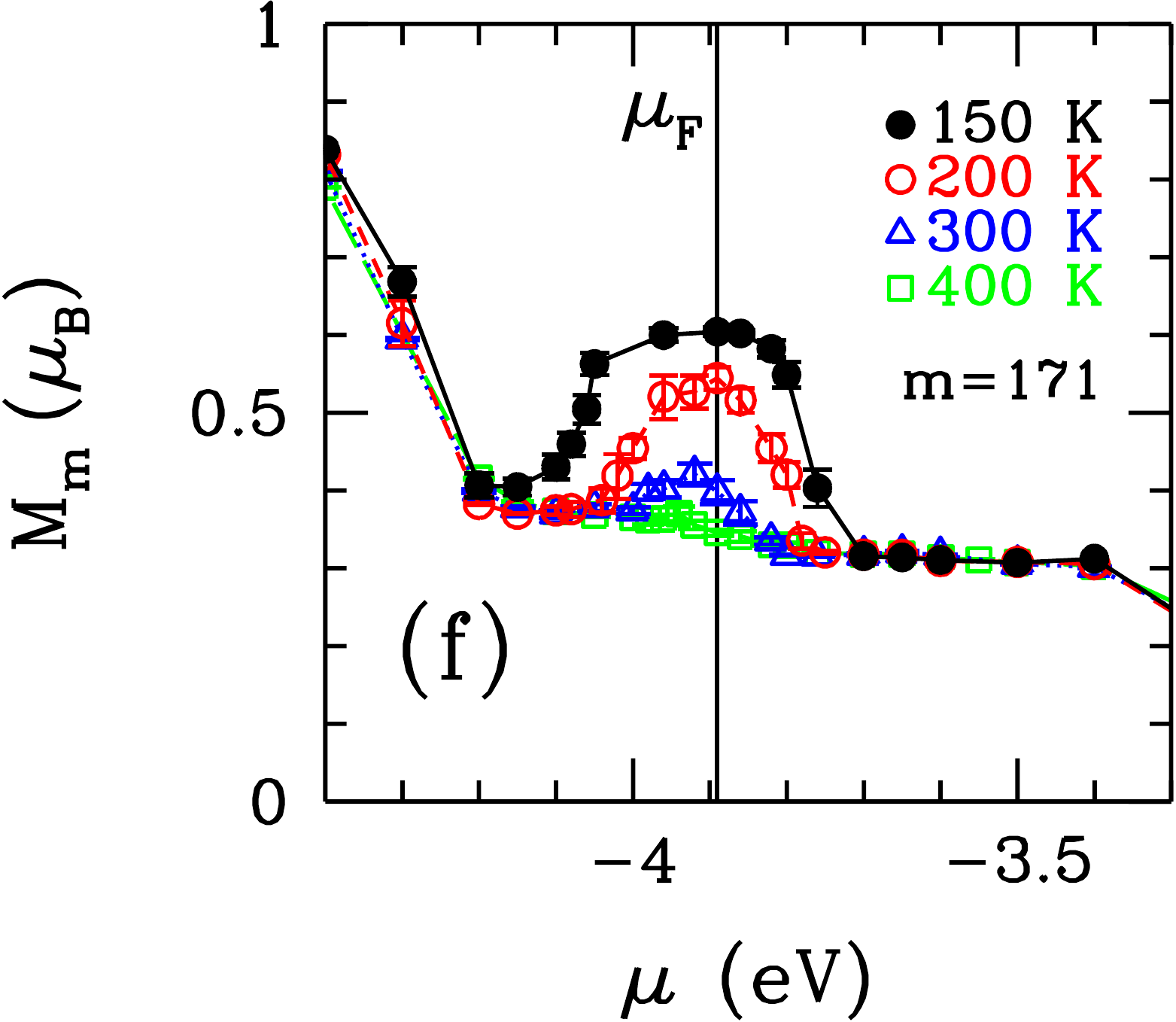}\vspace{1cm}}
\justify
(e) Electron number of the $m=171$'th host state 
$\langle n_m\rangle$ versus $\mu$. 
(f) Magnetic moment of the $m=171$'th host state 
$M_m$ versus $\mu$.
In these figures,
the black vertical line denotes the Fermi level $\mu_F$ 
at $T=150$ K. 
\end{figure}

\clearpage

\begin{figure}
\centering
\caption*{
{\bf Extended Data Fig. 3 $\vert$ Shift of the Fermi level 
with the temperature for oxy-heme} 
}
\vspace{1cm}
{\includegraphics[width=13cm]{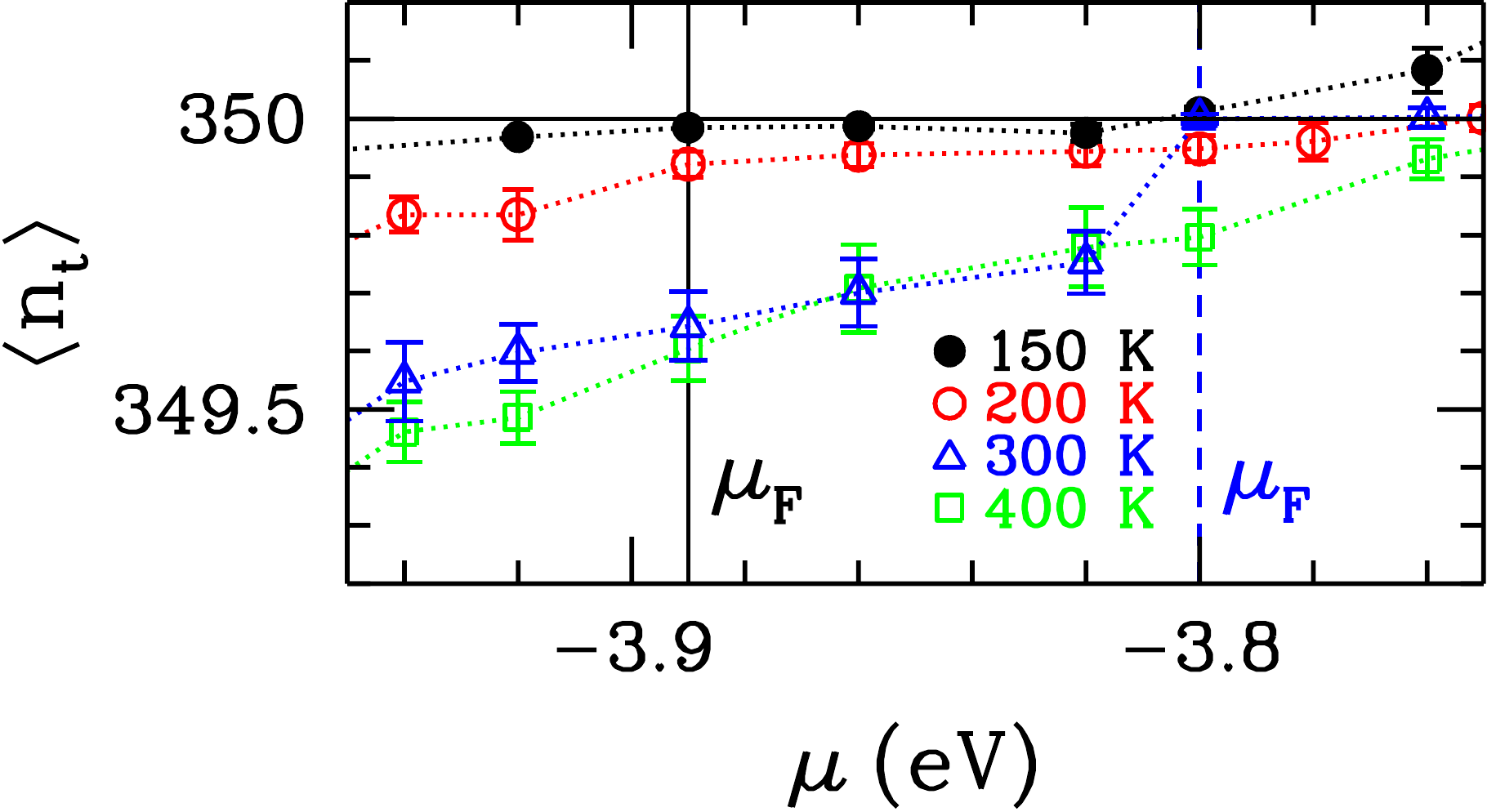}\hspace{0.0cm}}
\justify
Total electron number of the oxy-heme cluster
$\langle n_{\rm t}\rangle$ 
versus the chemical potential $\mu$ at various temperatures.
The charge-neutral oxy-heme cluster has 350 electrons. 
We see that, as the temperature is lowered, 
the Fermi level $\mu_{\rm F}$ shifts from $\approx -3.8$ eV at 300 K
to $\approx -3.9$ eV at 150 K.
\label{XDfig4}
\end{figure}

\begin{figure}
\centering
\caption*{
{\bf Extended Data Fig. 4 $\vert$ QMC finite-$\Delta\tau$ effects in the oxy-heme case} 
}
\vspace{1cm}
{\includegraphics[width=7.5cm]{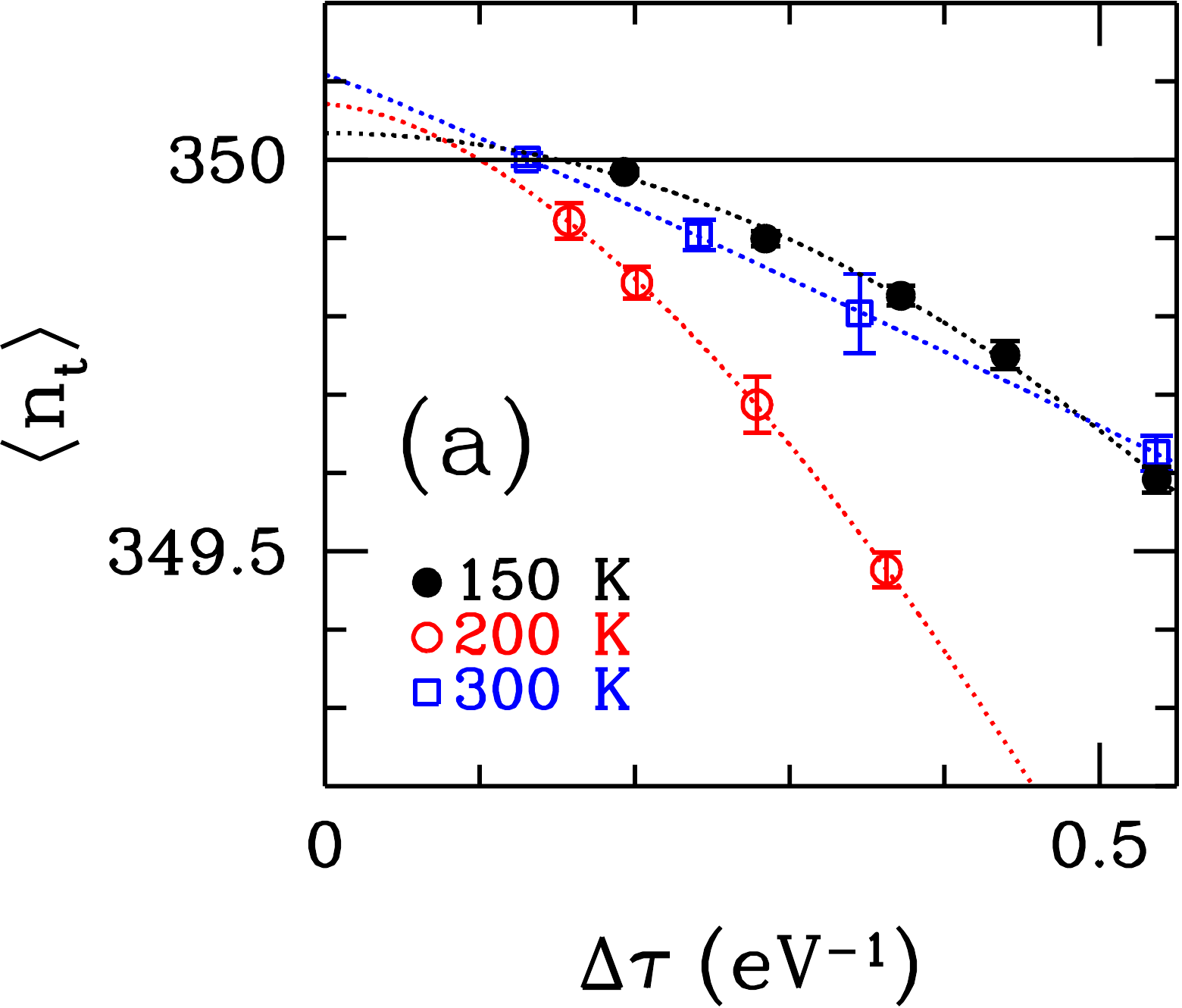}\hspace{0.5cm}}
{\includegraphics[width=7.5cm]{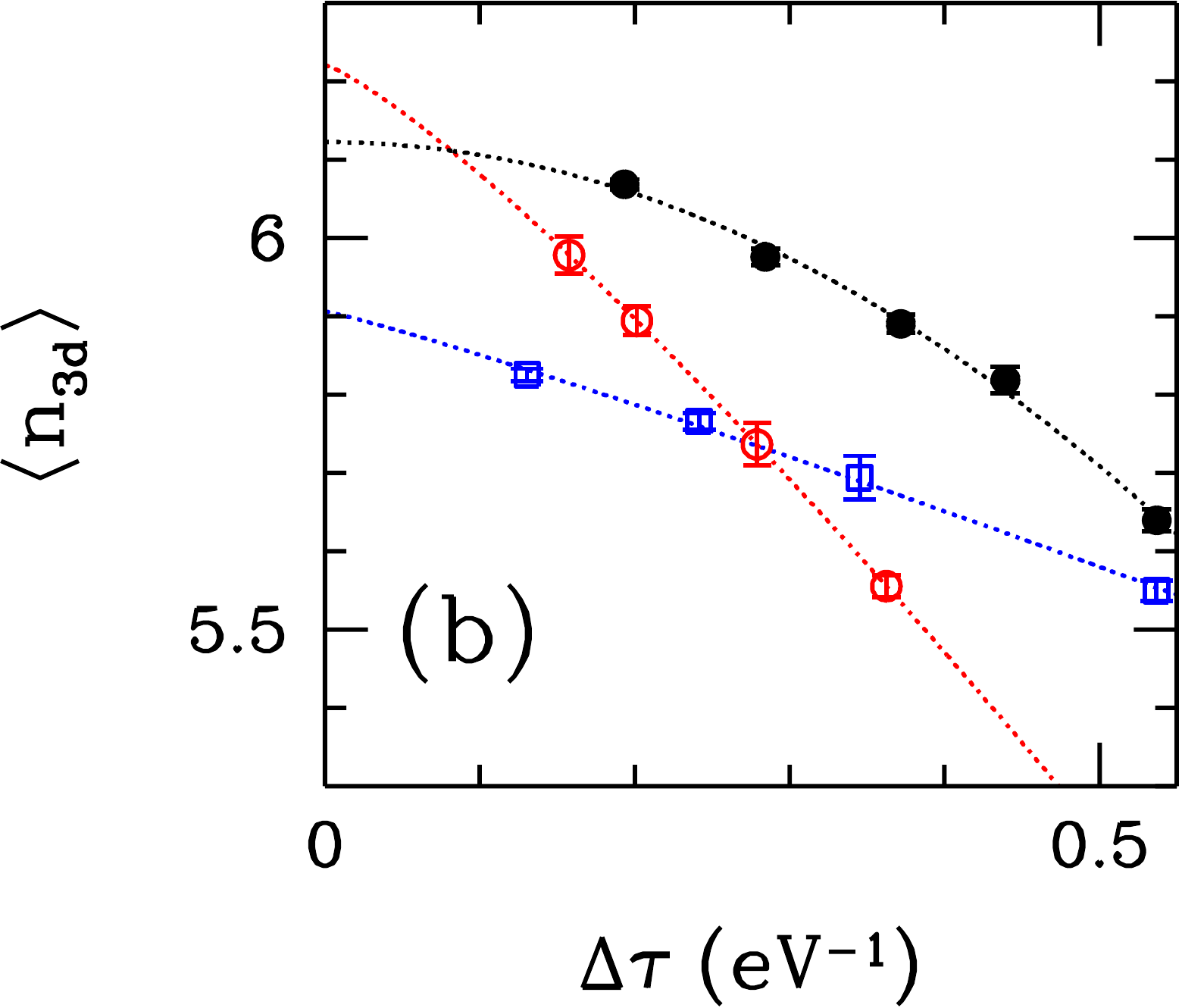}\vspace{1.0cm}}
{\includegraphics[width=7.5cm]{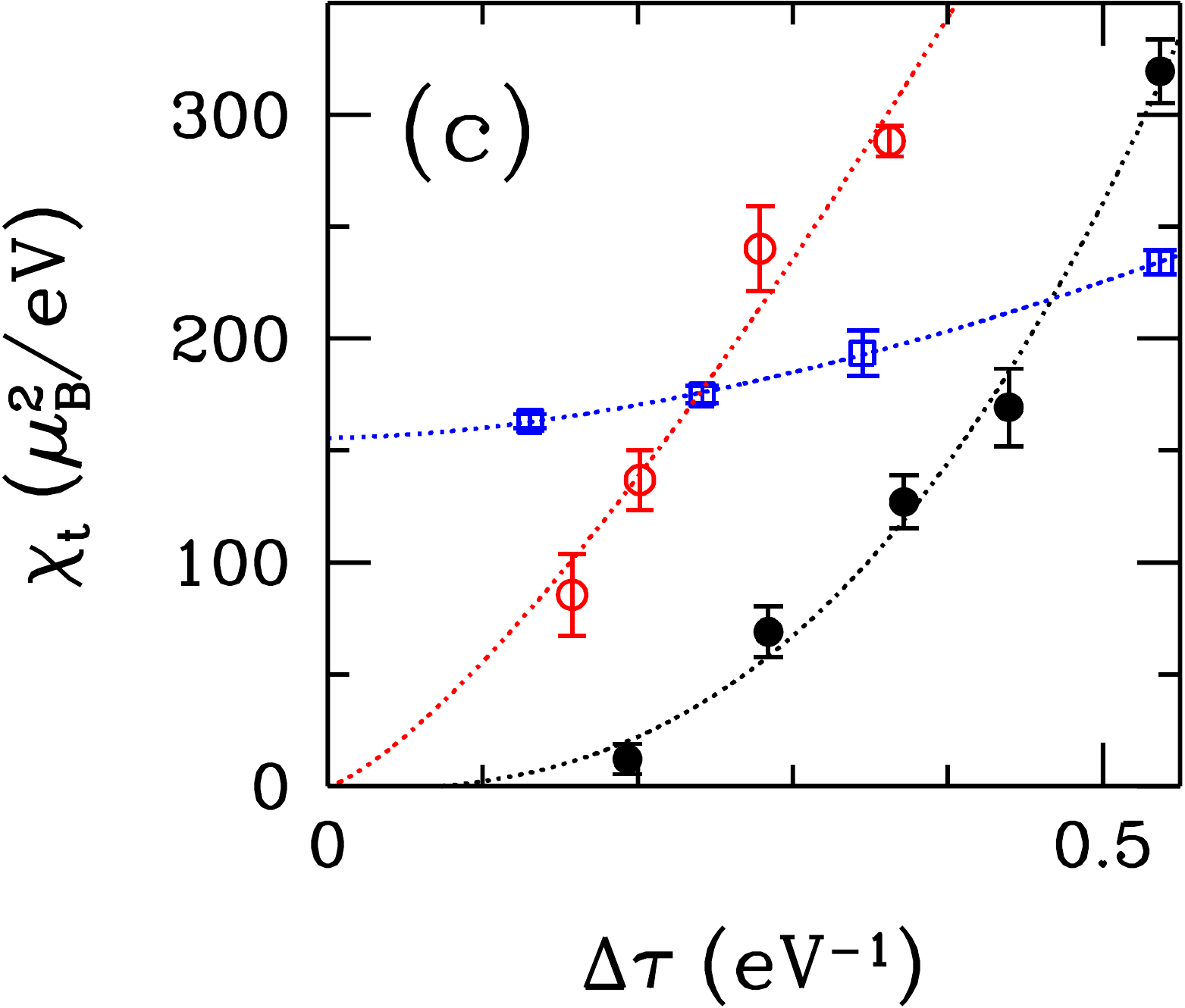}\hspace{0.5cm}}
{\includegraphics[width=7.5cm]{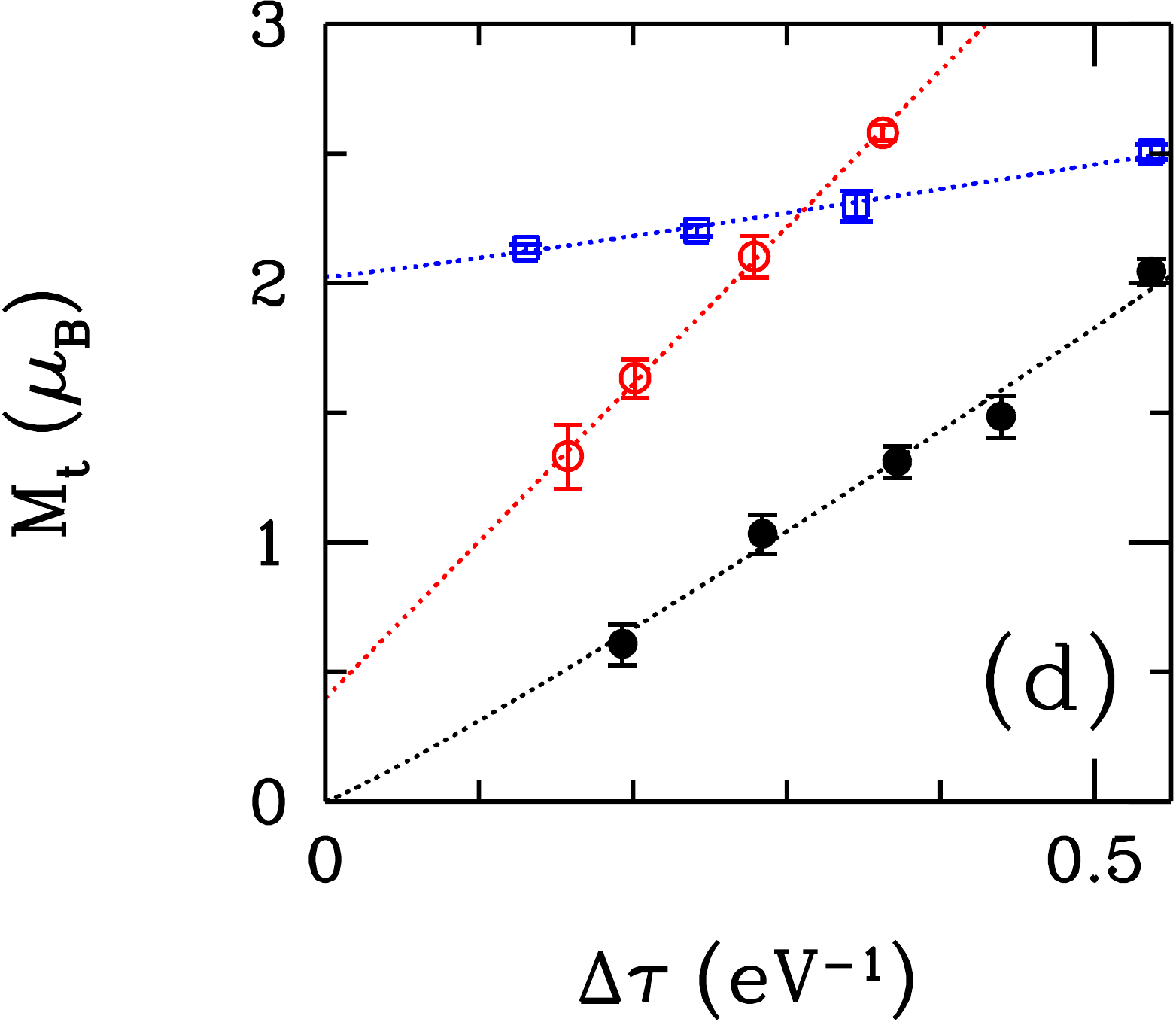}\vspace{1.0cm}}
\justify
(a) Total number of the electrons in the cluster 
$\langle n_{\rm t}\rangle$,
(b) the total electron number in the Fe($3d$) orbitals $\langle n_{3d}\rangle$,
(c) the total spin susceptibility $\chi_{\rm t}$,
and (d) the total magnetic moment $M_{\rm t}$
plotted as a function of $\Delta\tau$ 
for the various values of $T$ shown in (a). 
Here, 
$\Delta\tau$ is the Matsubara-time step used in the QMC simulations. 
The dotted curves are the least-squares fits to the QMC data. 
\label{XDfig4}
\end{figure}

\begin{figure}
\centering
\caption*{
{\bf Extended Data Fig. 5 $\vert$ Electronic transitions for the anomalous MCD spectrum
of deoxy-heme}
}
\vspace{1cm}
{\includegraphics[width=15.5cm]{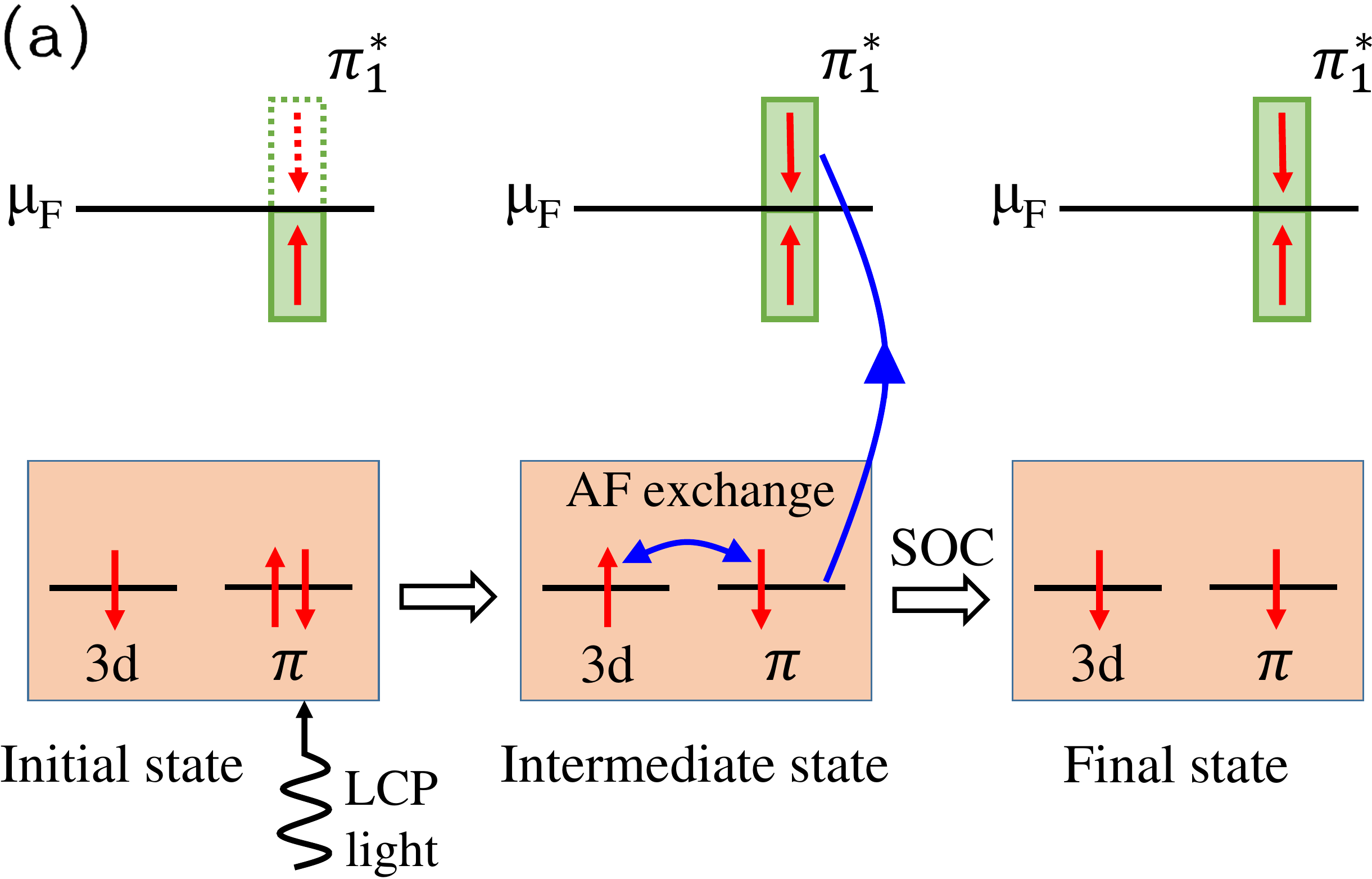}\vspace{1.5cm}} 
\justify
(a) Illustration of the initial, 
intermediate and the final states
in the $\pi\rightarrow \pi_1^*$ transition with LCP light absorption 
in MCD in the UV region.
These panels are
for an applied magnetic field pointing in the up direction, 
which is also along the direction of light propagation. 
\label{XDfig5}
\end{figure}

\clearpage

\begin{figure}
\centering
{\includegraphics[height=7.0cm]{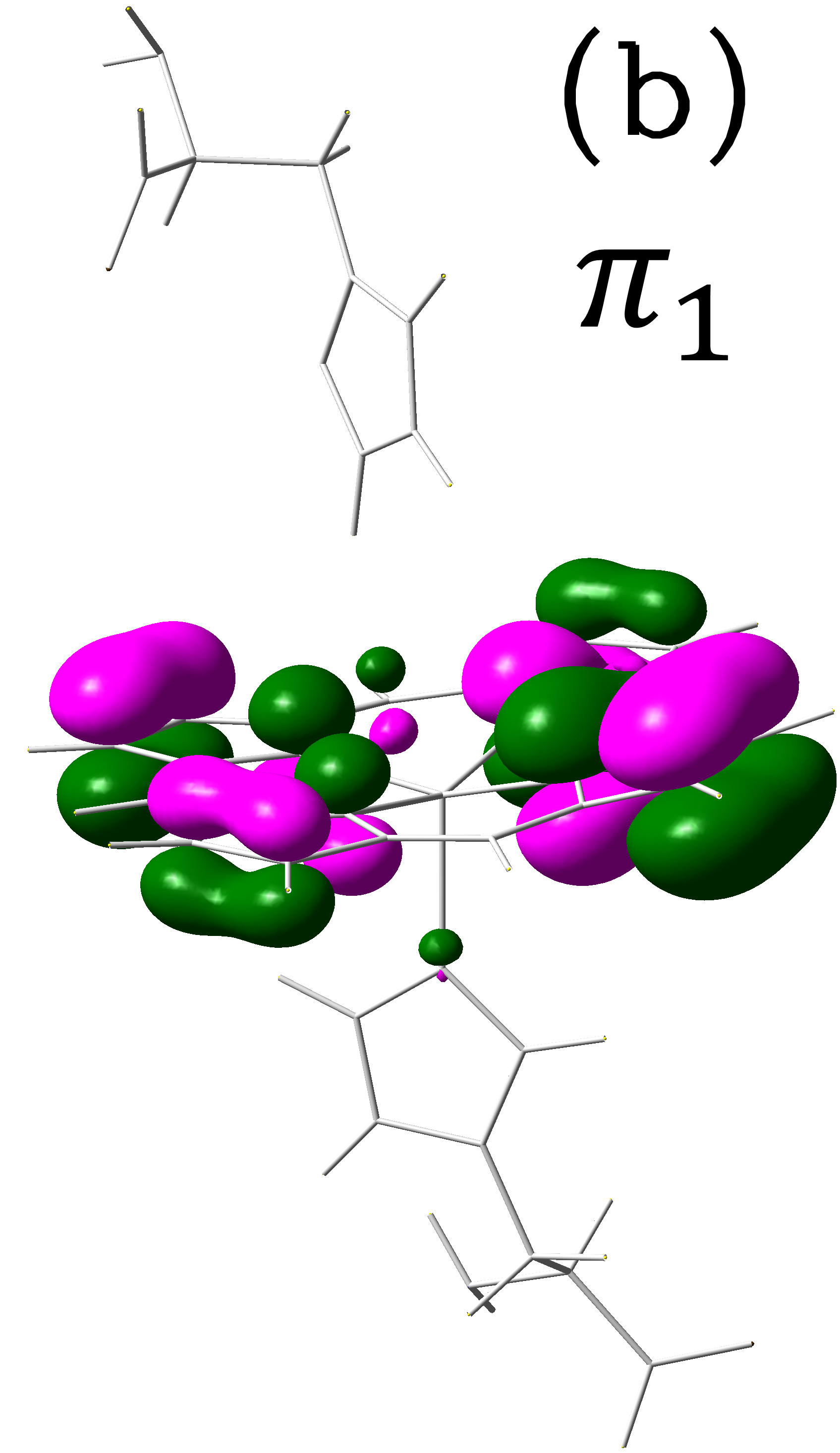}}\hspace{1.0cm}
{\includegraphics[height=7.0cm]{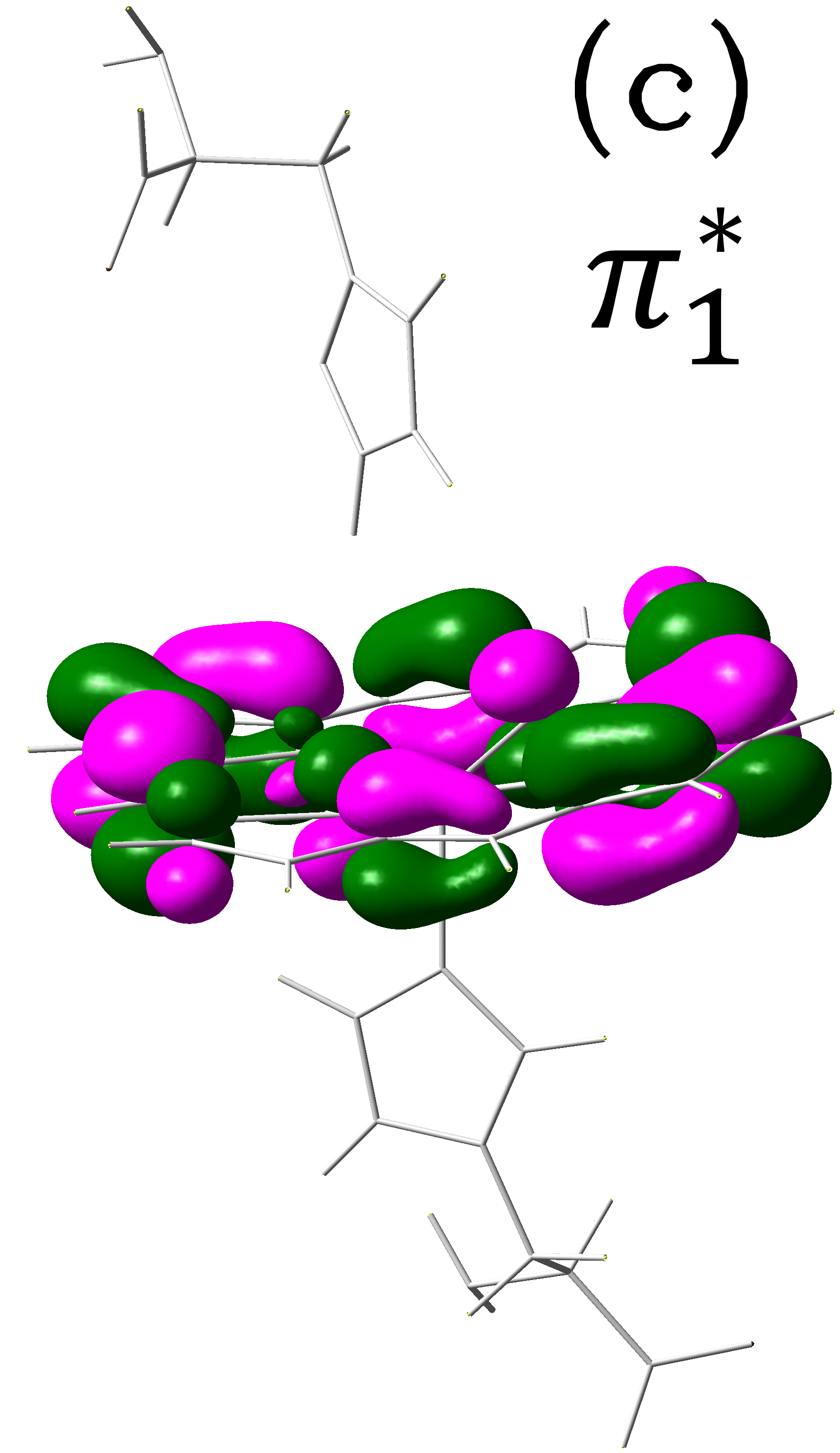}}\hspace{1.0cm}
{\includegraphics[height=7.0cm]{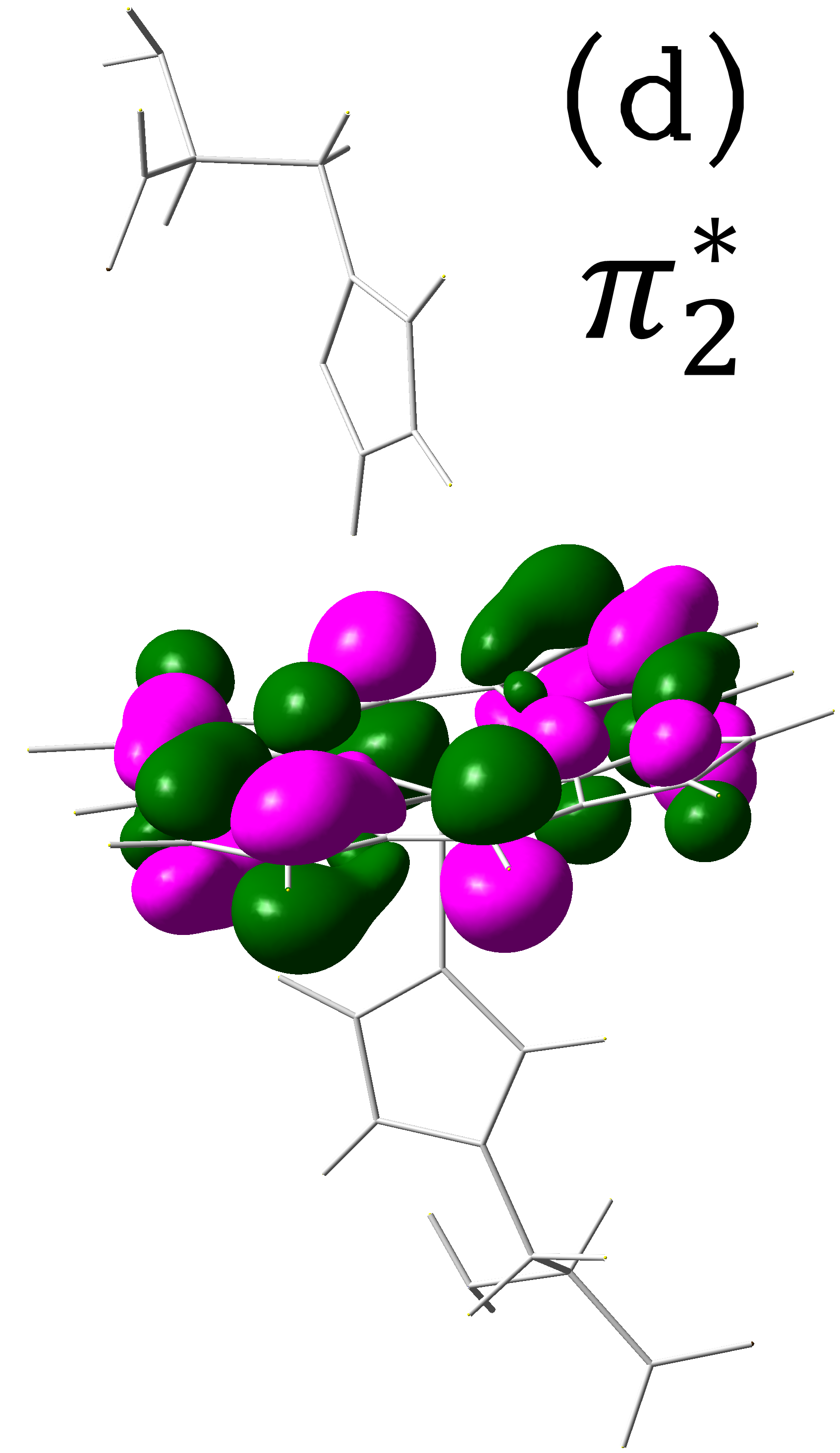}}\vspace{2.0cm}
{\includegraphics[width=15cm]{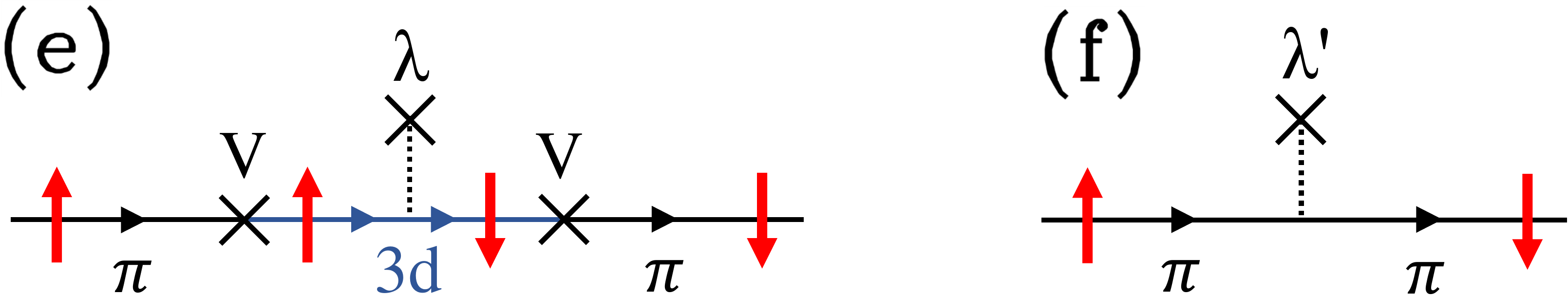}\vspace{1.0cm}}
\justify
(b)-(d) Illustration of the wavefunctions for the 
bonding $\pi_1$ and the antibonding $\pi_1^*$ and $\pi_2^*$ 
host states, respectively.
(e) Feynman diagram illustrating a hybridization process through 
which an electron in a bonding $\pi$ state gains an effective spin-orbit coupling. 
Here, an up-spin electron in the $\pi$ state can become an up-spin $3d$ electron
through hybridization. 
Because of the spin-orbit coupling at the Fe site, 
this up-spin $3d$ electron can now flip its spin down. 
Through hybridization for a second time, 
it then becomes a down-spin $\pi$ electron. 
This is the lowest-order diagram to the set of processes where 
the $\pi$ state gains an effective spin-orbit coupling.
Here, $V$ is the hybridization matrix element between the $\pi$ state 
and the Fe($3d$) orbitals,
and $\lambda$ is the spin-orbit coupling constant 
for the Fe($3d$) orbitals. 
(f) Because of processes as shown in (e), 
an electron in a $\pi$ state can gain an effective 
spin-orbit coupling $\lambda'$.
\label{XDfig5-2}
\end{figure}

\begin{figure}
\centering
\caption*{
{\bf Extended Data Fig. 6 $\vert$ Electronic properties of the $\pi_1^*$ 
and $\pi_2^*$ host states located near the Fermi level in deoxy-heme}
}
\vspace{1cm}
{\includegraphics[width=7.5cm]{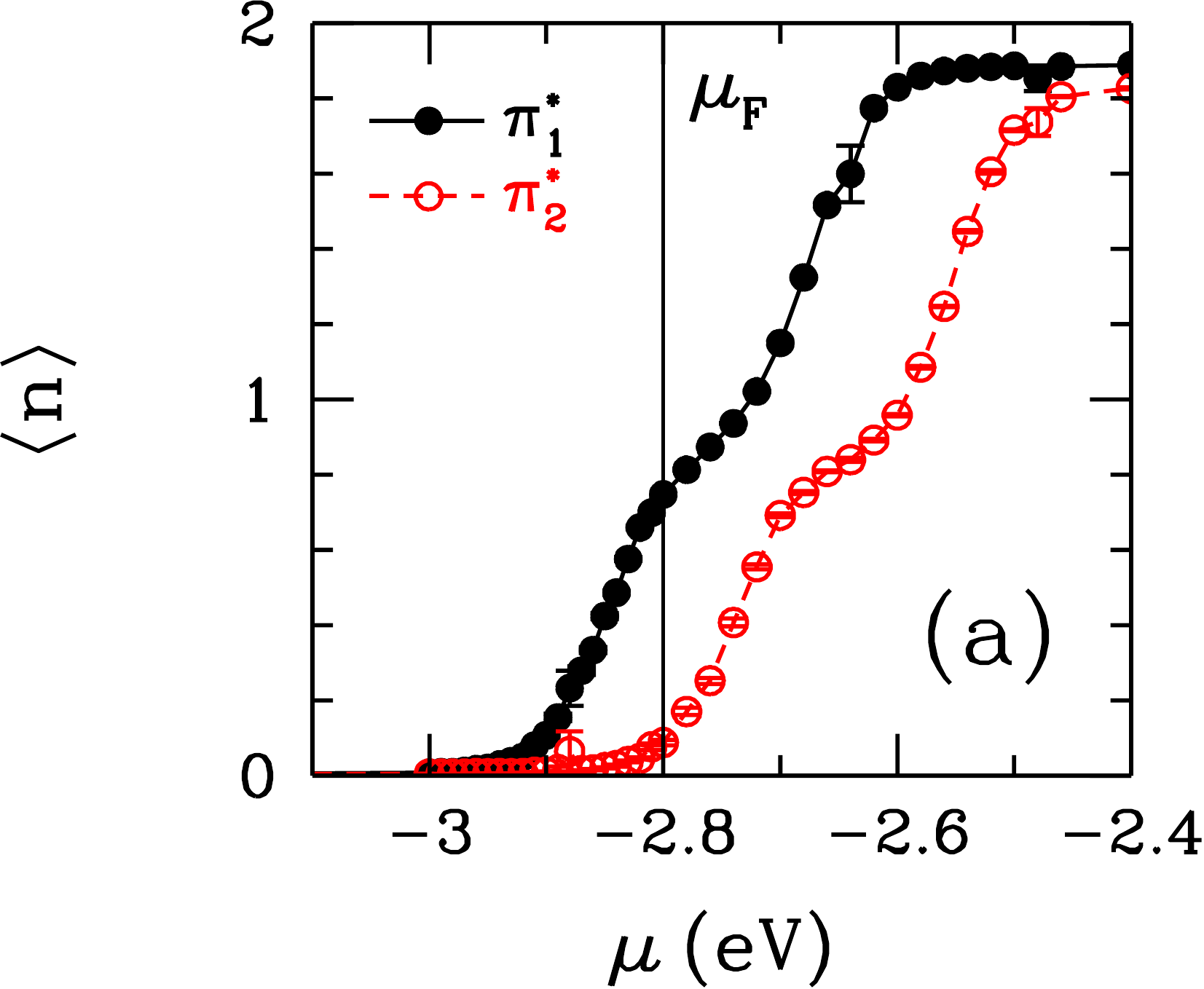}\hspace{0.5cm}}
{\includegraphics[width=7.5cm]{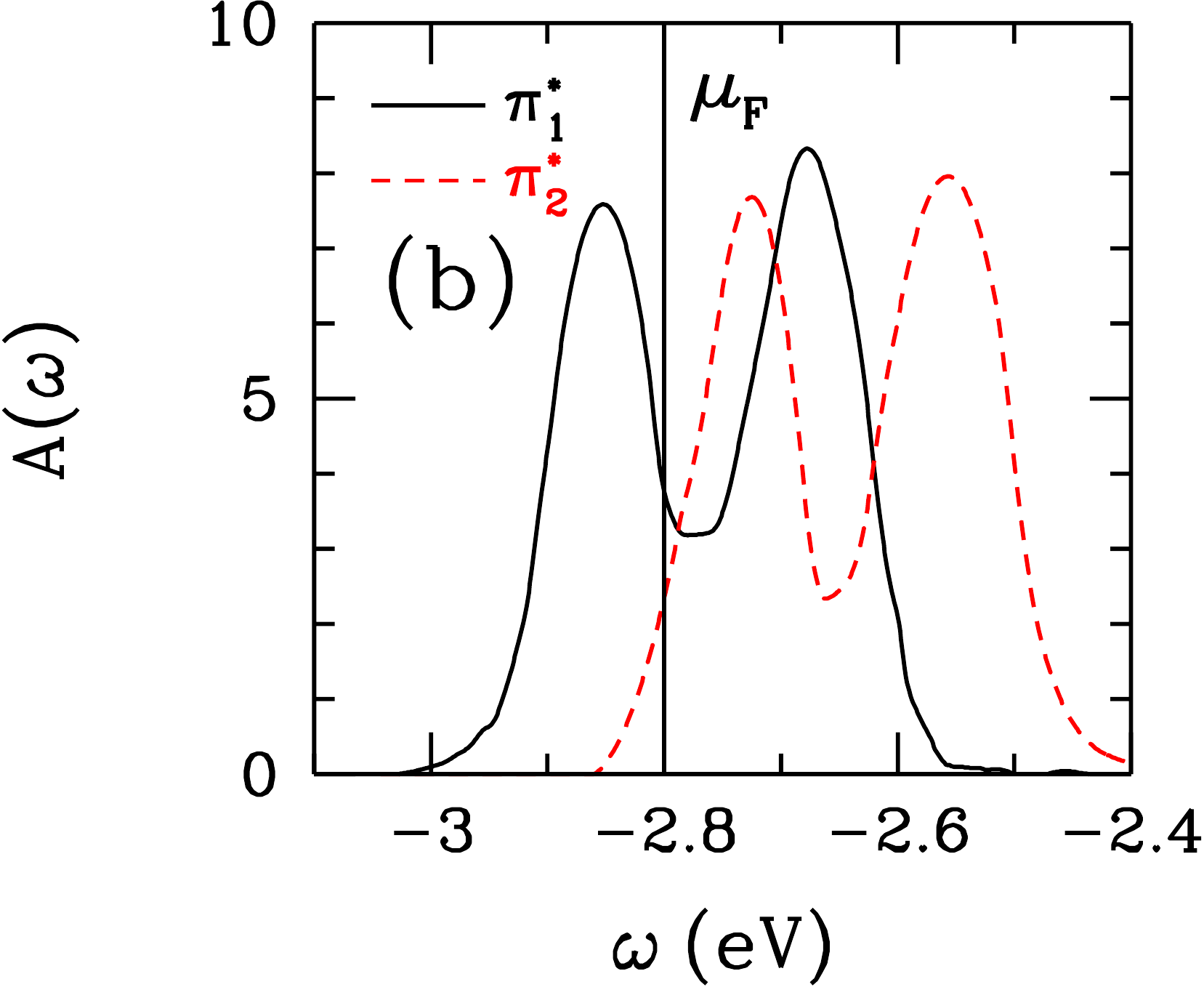}\vspace{1cm}}
{\includegraphics[width=7.5cm]{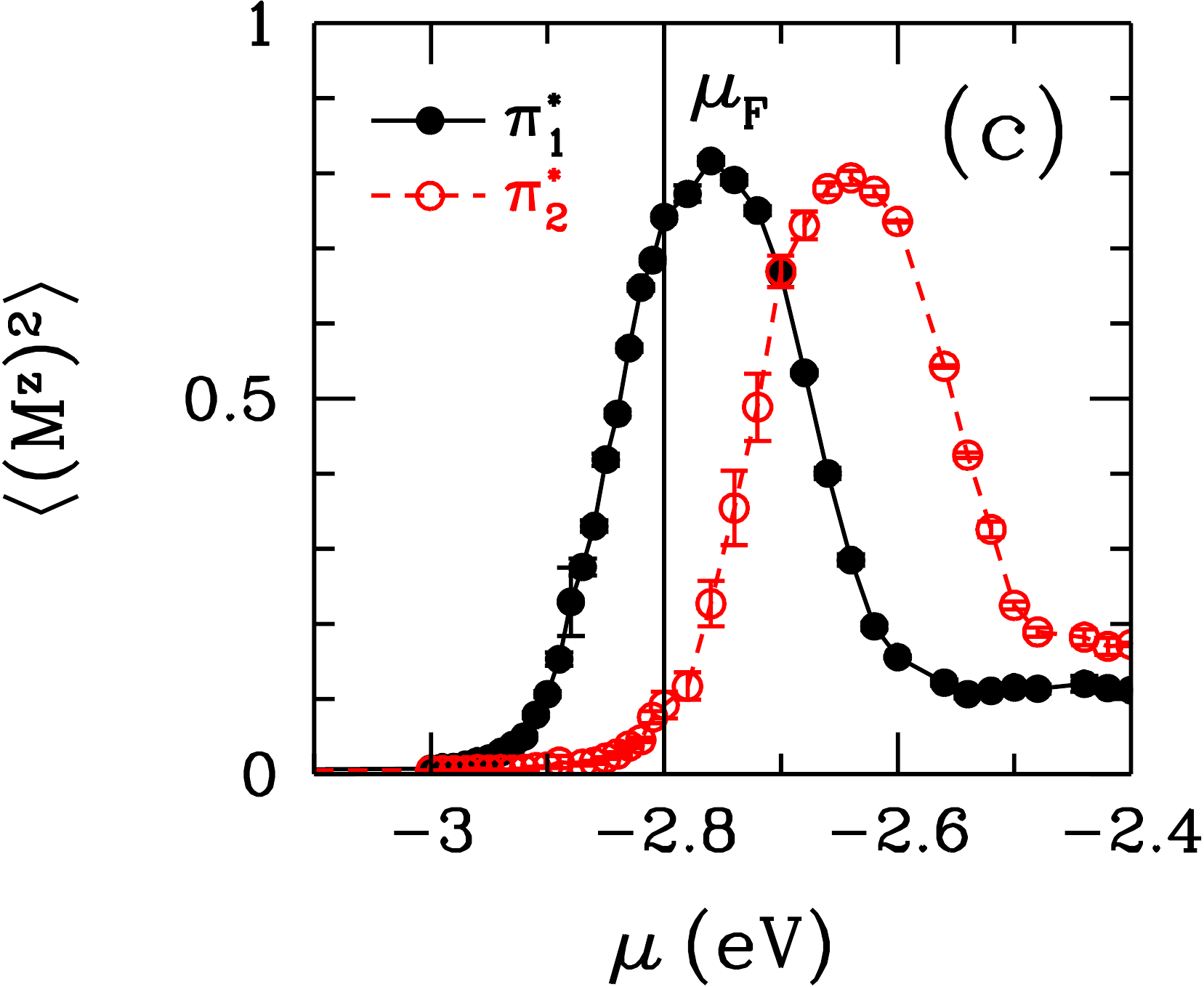}\hspace{0.5cm}}
{\includegraphics[width=7.5cm]{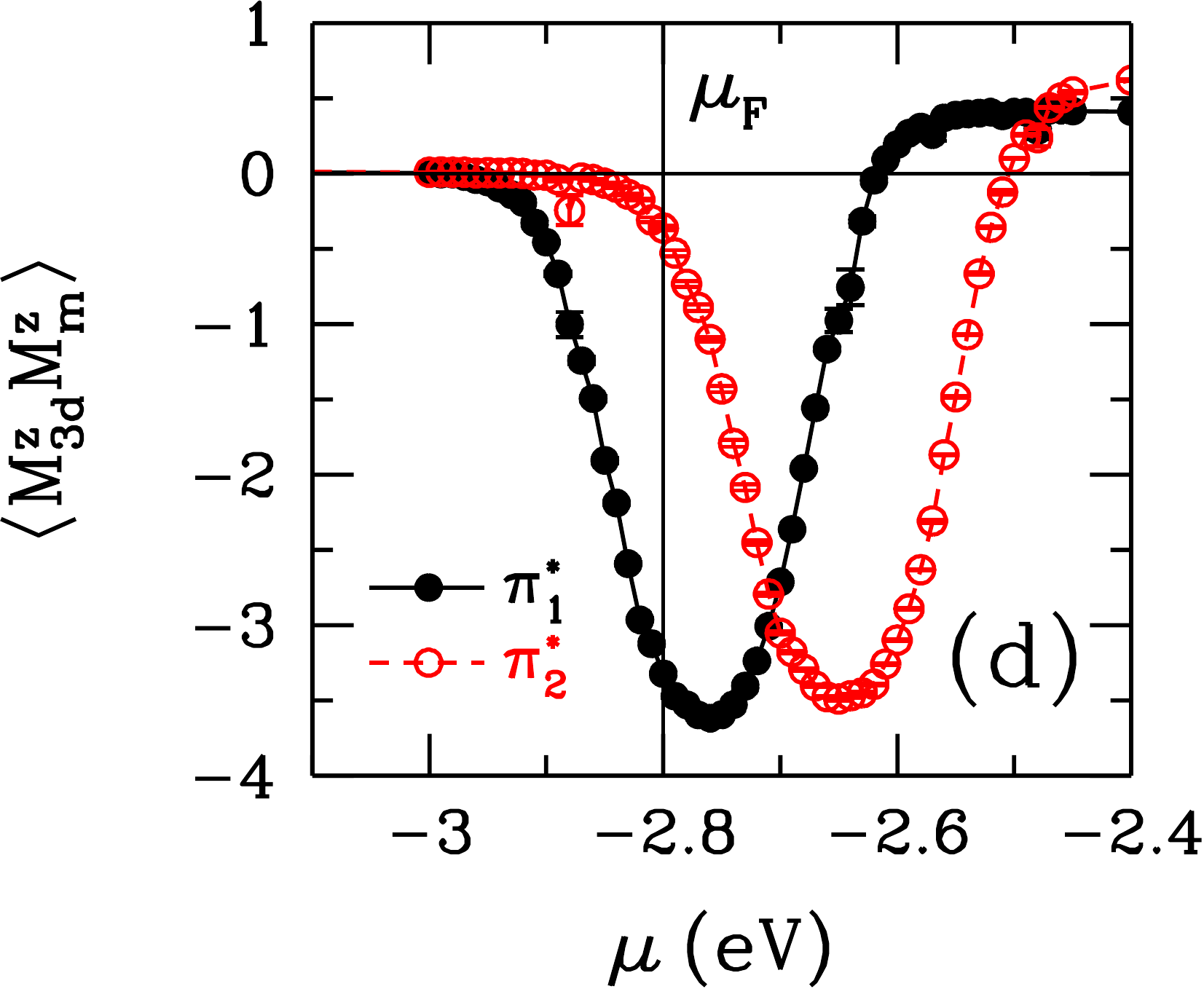}\vspace{1cm}}
\justify
(a) Electron occupation $\langle n\rangle$ 
of the $\pi_1^*$ and $\pi_2^*$ host states plotted 
as a function of the chemical potential $\mu$. 
(b) Single-particle spectral weight $A(\omega)$ versus $\omega$ 
for the $\pi_1^*$ and $\pi_2^*$ states
obtained from the $\langle n\rangle$ versus $\mu$ results shown in (a). 
(c) Square of the magnetic moment of the $\pi_1^*$ and $\pi_2^*$ states 
versus $\mu$.
(d) Magnetization correlation function of the Fe($3d$) magnetic moment 
with the moments of the $\pi_1^*$ and $\pi_2^*$ states 
$\langle M^z_{3d} M^z_m \rangle$ 
versus $\mu$.
In these figures, 
the black vertical line denotes the Fermi level $\mu_F$, 
and the results are presented for $T=300$ K. 
\label{XDfig6}
\end{figure}

\begin{figure}
\centering
\caption*{
{\bf Extended Data Fig. 7 $\vert$ Experimental and calculated results 
on the optical absorption in UV region for deoxy-heme}
}
\vspace{0cm}
{\includegraphics[width=13cm]{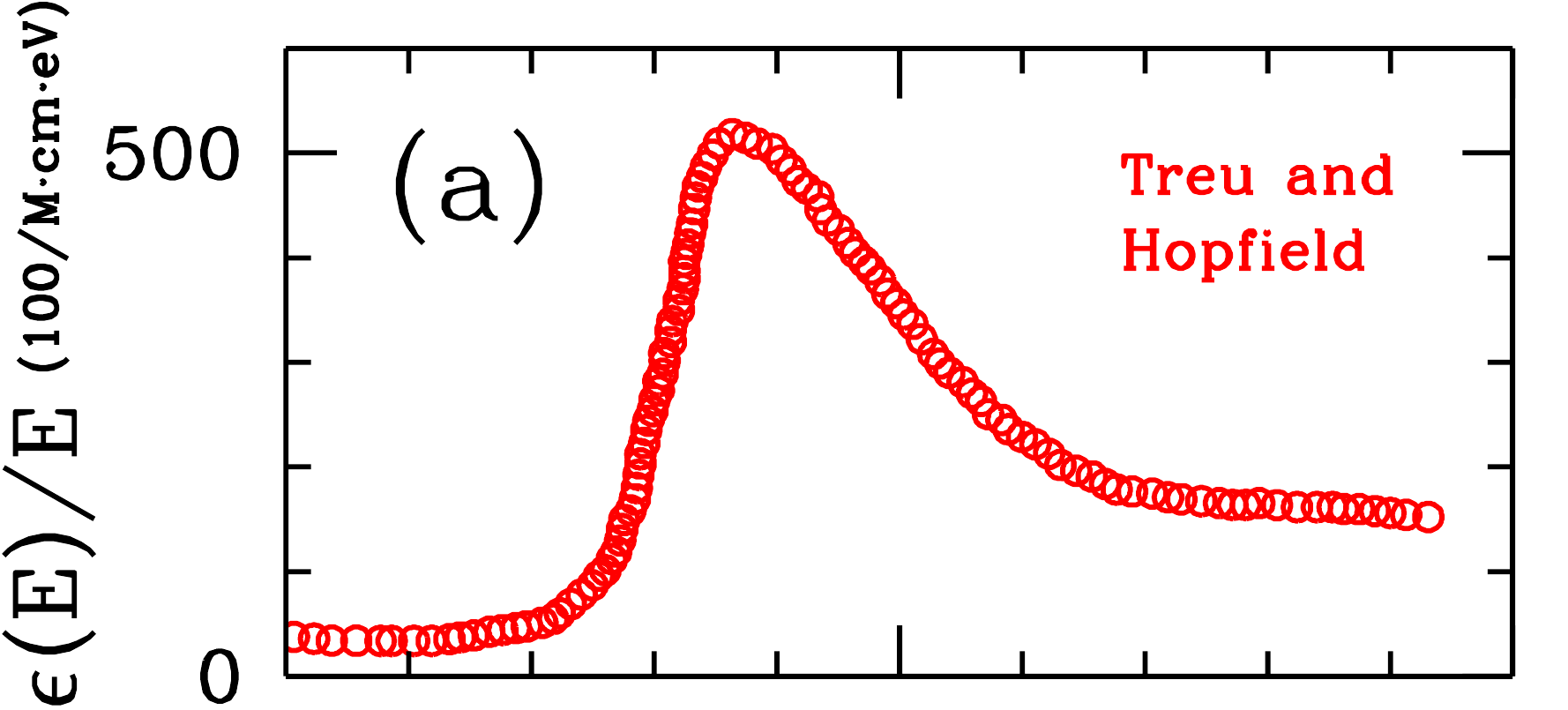}\vspace{0.5cm}}
{\includegraphics[width=13cm]{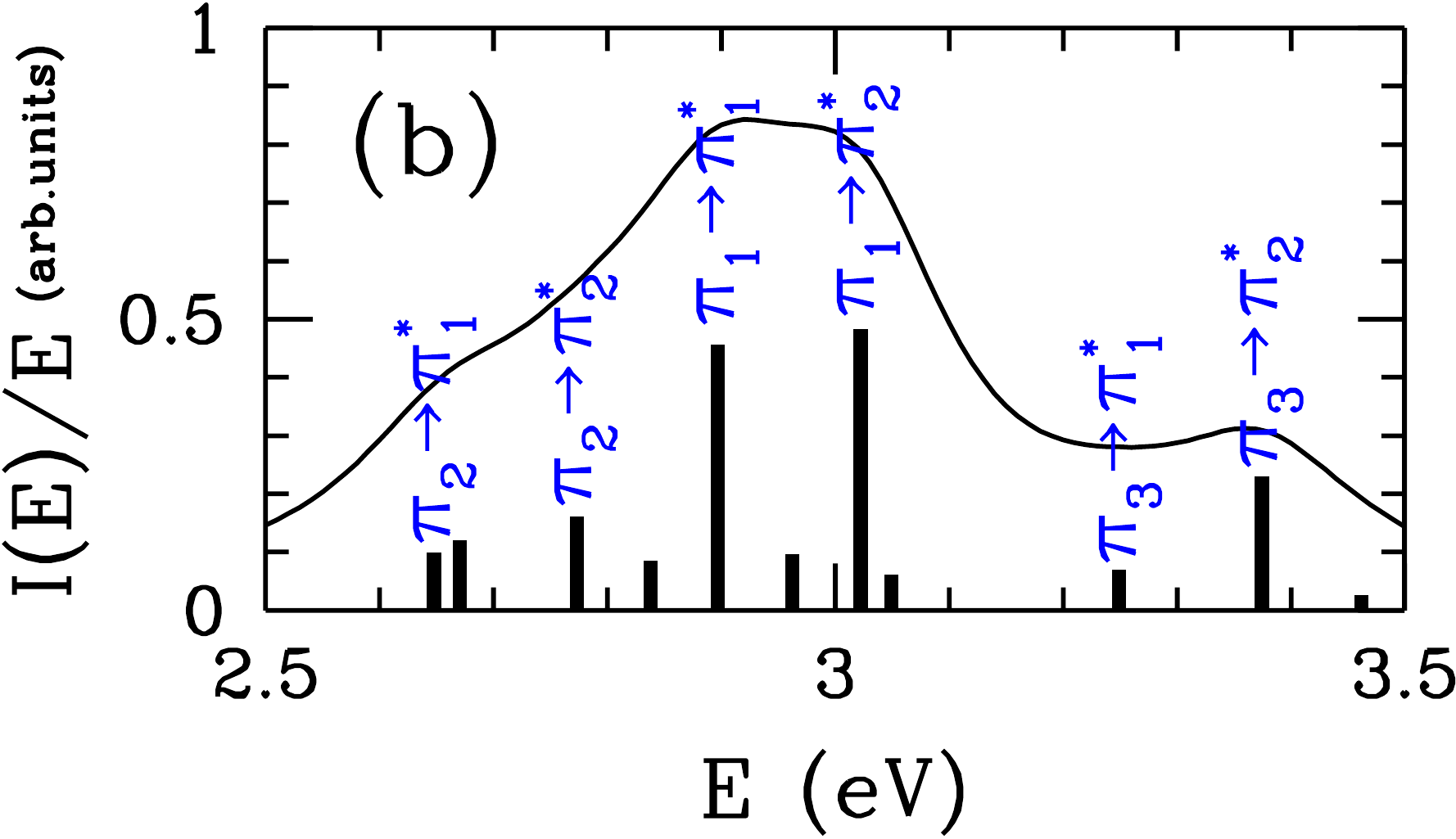}}
\justify
(a) Experimental data on the frequency dependence of the optical absorption 
normalized by energy
$\varepsilon(E)/E$
for deoxy-HbA by Treu and Hopfield \cite{Treu}.
(b) Calculated optical absorption normalized by energy $I(E)/E$ 
for deoxy-heme.
The black bars denote the weights of the various $\pi\rightarrow \pi^*$ 
transitions. 
Here, we have labelled the $\pi$ states
which give the leading contributions
as $\pi_1$, $\pi_2$ and $\pi_3$.
We have also indicated which particular $\pi\rightarrow \pi^*$ 
transitions the bars correspond to. 
The black curve was obtained by artificially broadening 
the delta functions by 0.1 eV. 
\label{XDfig7}
\end{figure}

\begin{figure}
\centering
\caption*{
{\bf Extended Data Fig. 8 $\vert$ MCD spectrum
due to $\pi_1 \rightarrow (\pi_1^*,\pi_2^*)$ optical transitions}
}
\vspace{1cm}
{\includegraphics[width=13cm]{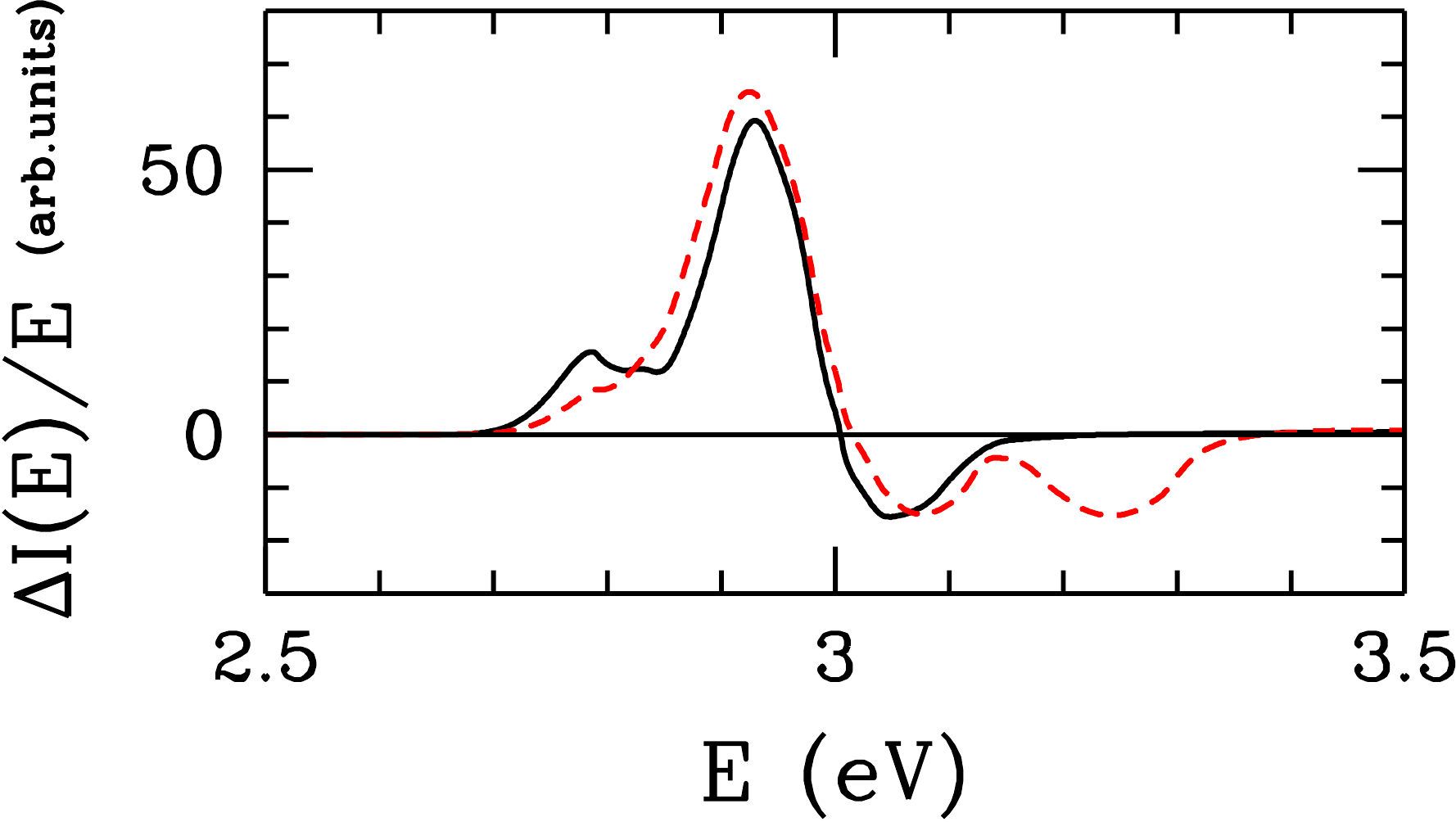}}
\justify
Here, the black curve denotes the MCD spectrum for when only 
the $\pi_1\rightarrow \pi_1^*$ and 
$\pi_1\rightarrow \pi_2^*$ transitions are taken into account. 
The red-dashed curve denotes the same in the case where 
the $\pi_2^*$ state is artificially shifted by 0.2 eV to higher energies 
so that there is no overlap with the $\pi_1^*$ state. 
This yields the best agreement with the experimental line shape seen in Fig. 4(d). 
\label{XDfig8}
\end{figure}

\clearpage

{\bf{SUPPLEMENTARY INFORMATION} }\\

{\bf Calculation of the Anderson model parameters with DFT for heme clusters} \\
We obtain the one-electron parameters $\varepsilon_{m}$, 
$\varepsilon_{d\nu}$ and 
$V_{m \nu}$ of the effective Anderson model 
by making use of the DFT method. 
These calculations are performed for a reduced cluster 
with molecular coordinates obtained from the Protein Data Bank.
In this calculation, 
instead of the atomic orbitals, 
we use the 
natural atomic orbitals (NAO's) \cite{Reed}, 
which form an orthonormal basis. 
First, we express the Fock matrix in the NAO basis. 
We take the Fe($3d_{\nu}$) NAO's as the impurity orbitals and 
their energy levels as $\varepsilon_{d\nu}$'s in the Anderson Hamiltonian. 
Diagonalizing the remaining part of the Fock matrix, 
we obtain the host eigenstates $|u_m\rangle$ 
and their energy levels $\varepsilon_{m}$
and
the hybridization matrix elements $V_{m\nu}$.
This procedure is explained in more detail in Ref. \cite{Kandemir,Mayda}.
The DFT calculations are carried out by using 
the Gaussian program \cite{Gaussian} with 
the BP86 energy functional \cite{Becke,Perdew} 
and the 6-31G basis set with $483$ 
basis functions for the deoxy-heme cluster and 
$501$ basis functions for the oxy-heme cluster. 
We use $\varepsilon_m$, 
$\varepsilon_{d\nu}$ and $V_{m\nu}$ 
determined this way as input parameters for 
the QMC simulations. 

{\bf Truncated heme clusters and the finite-size effects}\\
The deoxy-HbA molecule seen in Fig. 1(a) 
contains four inequivalent heme groups, 
$\alpha_1-\alpha_2-\beta_1-\beta_2$.
The nearest-neighbor Fe-Fe distance varies between 34 {\AA} 
and 39.5 {\AA}.
Since this molecule consists of about 9700 atoms, 
we performed our calculations for a reduced heme cluster 
obtained from the $\alpha_1$ group, which is shown in Fig. 1(b). 
This cluster
contains the porphyrin ring with Fe at the center,
located below porphyrin is the distal histidine with the imidazole part,
and also located at the top is the proximal histidine. 
It is thought that the proximal histidine is necessary for 
the stability of O$_2$ binding \cite{Birukou}. 
In obtaining this truncated cluster, 
we have replaced the methyl, vinyl and propionate 
side groups of porphyrin with hydrogen atoms. 
We have determined the coordinates of these substituted hydrogens by DFT optimization. 
The deoxy-heme cluster obtained this way
consists of 75 atoms and 334 electrons 
(C$_{32}$H$_{30}$FeN$_{10}$O$_{2}$). 
In order to obtain the oxy-heme cluster, 
we have again started with the molecular structure of oxy-HbA 
determined by X-ray measurements from the Protein Data Bank
(Keyword: 2DN1).
From the $\alpha_1$ heme group we have obtained 
the cluster seen in Fig. 1(c) with 77 atoms and 350 electrons 
(C$_{32}$H$_{30}$FeN$_{10}$O$_{4}$).

In obtaining the energy spectrum of the cluster with DFT,
it is necessary to take into account the finite size effects
arising from the boundary of the cluster. 
Even though the HbA molecule contains about 9700 atoms, 
we retain only 75 atoms for the deoxy-heme cluster shown in Fig. 1(b).
In this case we find that a host state which is localized 
on the oxygen and carbon sites at the boundary of the cluster
(at the lower edge of the cluster shown in Fig. 1(b))
has an energy close to the Fermi level. 
However, when we use larger clusters containing 87 or 96 atoms
so that the distal histidine part contains more sites
and the oxygen site is not close to the boundary,
we find that this host state arising from the boundary 
moves away from the Fermi level to higher energies. 
Hence,
for the 75 site cluster we have by hand removed the boundary host state 
in order to control the finite size effects. 

{\bf QMC measurements} \\
In the DFT+QMC approach, 
the Coulomb interactions at the Fe($3d_{\nu}$) orbitals are taken into 
account by both the DFT and the QMC techniques. 
In order to prevent this double counting, an 
orbital-dependent 
double-counting term $\mu_{\nu}^{\rm {DC}}$ 
defined as 
\begin{eqnarray}
\mu^{\mathrm {DC}}_{\nu} = \frac{1}{2} U \, n^0_{d\nu} 
+ \frac{1}{2} (U'+U'') \sum_{\nu' \neq \nu} \,
n^0_{d \nu'}
\end{eqnarray}
is substracted from the bare Fe($3d_{\nu}$) energy levels,
$\varepsilon_{d\nu} 
\rightarrow 
\varepsilon_{d\nu} - \mu_{\nu}^{\rm {DC}}$. 
Here, $n^0_{d\nu}$ 
is the electron number in the Fe($3d_{\nu}$) NAO's obtained by the DFT calculations.

By using QMC simulations we calculate the expectation values
of various operators to study the electronic properties 
of the effective Anderson model for heme.
In particular, 
we calculate the distribution of the magnetic moments in the cluster, 
the correlations among these moments,
the magnetic susceptibilities and 
in addition the charge distribution throughout the cluster. 
The electron occupation number of the Fe($3d_{\nu}$) orbitals is obtained from
\begin{equation}
\langle n_{\nu}\rangle = 
\sum_{\sigma} 
\langle d_{\nu\sigma}^{\dagger} d_{\nu\sigma} \rangle,
\end{equation}
where the expectation value of an operator $A$ is defined by
\begin{equation}
\langle A\rangle = \frac{1}{Z} {\rm Tr} \, e^{-\beta H} A
\end{equation}
with $Z= {\rm Tr} \, e^{-\beta H}$ the partition function.
In addition, we calculate the 
effective magnetic moments $M_{\nu}^{\rm eff}$ of the Fe($3d_{\nu}$) orbitals from 
\begin{equation}
M_{\nu}^{\rm eff} =
\sqrt{\langle (M_{\nu}^{z})^{2} \rangle},
\end{equation}
where the longitudinal magnetization operator for the Fe($3d_{\nu}$) orbital 
is defined in units of $\mu_{\rm B}$ as
\begin{equation}
M_{\nu}^z = d_{\nu\uparrow}^{\dagger} d_{\nu\uparrow} -
d_{\nu\downarrow}^{\dagger} d_{\nu\downarrow}. 
\end{equation}
We calculate the total Fe($3d$) spin susceptibility from 
\begin{equation}
\chi_{3d} = \int_0^{\beta} d \tau \,
\langle M_{3d}^z(\tau) \, M_{3d}^z(0) \rangle
\end{equation}
where the total Fe($3d$) magnetization operator is
$M_{3d}^z = \sum_{\nu} M_{\nu}^z$
with the Matsubara-time evolution 
$M^z_{3d}(\tau) = \exp(H\tau) \, M^z_{3d} \, \exp(-H\tau)$.
Similarly, the total spin susceptibility of the cluster is obtained from 
\begin{equation}
\chi_{\rm t} = \int_0^{\beta} d\tau \,
\langle M_{\rm t}^z(\tau) \, M_{\rm t}^z(0) \rangle
\label{chit}
\end{equation}
where the total magnetization operator of the cluster is
$M_{\rm t}^z = M_{3d}^z + M_{\rm h}^z$
and the total magnetization operator of the host is
$M_{\rm h}^z = \sum_m M_m^z$ with 
\begin{equation}
M_{m}^z = c_{m\uparrow}^{\dagger} c_{m\uparrow} -
c_{m\downarrow}^{\dagger} c_{m\downarrow}
\end{equation}
in units of $\mu_{\rm B}$. 
In evaluating the integrand in Eq. (7),
we include the correlations among the Fe($3d_{\nu}$) orbitals, 
and the Fe($3d$)-host correlations 
in addition to the intra-orbital host correlations as seen in
\begin{equation}
\langle M_{\rm t}^z(\tau) \, M_{\rm t}^z(0) \rangle =
\sum_{\nu,\nu'} \langle M_{\nu}^z(\tau) \, M_{\nu'}^z(0) \rangle +
2 \sum_{\nu,m} \langle M_{\nu}^z(\tau) \, M_{m}^z(0) \rangle +
\sum_{m} \langle M_{m}^z(\tau) \, M_{m}^z(0) \rangle.
\end{equation}
However, 
we do not include the contribution coming from the inter-orbital host correlations
\begin{equation}
\sum_{m,m'\neq m} \langle M_{m}^z(\tau) \, M_{m'}^z(0) \rangle,
\end{equation}
because its effect on $\chi_{\rm t}$ is  
negligible compared to the other terms. 
In the oxy-heme case, 
we estimate its contribution to $\chi_{\rm t}$ to be less than 
$\approx 4\, \mu_{\rm B}^2/{\rm eV}$.
 
{\bf Calculation of the magnetic-moment density} \\
Here, we describe how we calculate the magnetic-moment density $M({\bf r})$
for the heme clusters. 
The effective magnetic moment of the $m$'th host state is obtained from
\begin{eqnarray}
M_m^{\rm eff} = \sqrt{ \langle (M_m^z)^2 \rangle }.
\end{eqnarray}
While constructing the effective Anderson Hamiltonian for the heme cluster
from the Fock matrix,
we have obtained the expansion of the host states 
in terms of the atomic orbitals,
\begin{eqnarray}
c_{m\sigma} = \sum_i \, D_{m,i} \, \tilde{c}_{i\sigma},
\end{eqnarray}
where $\tilde{c}_{i\sigma}$ destroys an electron 
with spin $\sigma$ in the $i$'th atomic orbital. 
Hence, the magnetization operator of the $m$'th host state can be written as
\begin{eqnarray}
M_m^z = \sum_{i,j} D^*_{m,i} D_{m,j} 
( \tilde{c}^{\dagger}_{i\uparrow} \tilde{c}_{j\uparrow} - 
\tilde{c}^{\dagger}_{i\downarrow} \tilde{c}_{j\downarrow} ).
\end{eqnarray}

In order to obtain a simple illustration of the magnetization density,
we neglect the correlations for when $i\neq j$,
and define the following approximate expression 
as the magnetic-moment density for the host states
\begin{eqnarray}
M({\bf r}) \approx \sum_{m,i} \, M_m^{\rm eff} \, |D_{m,i}|^2 \, 
\delta({\bf r}-{\bf r}_i),
\end{eqnarray}
where ${\bf r}_i$ is the coordinate of the $i$'th atomic orbital.
Using this expression, 
we show in Fig. 2 the Fe($3d$) and the host magnetic-moment
density in the basis of the atomic orbitals as a bubble graph. 
We use the sign of the Fe($3d$)-host correlation function
$\langle M_{3d}^z M_m^z\rangle$ to determine the sign of $M({\bf r})$. 
In obtaining $M({\bf r})$ from Eq. (14),
we have only included the contributions from host states
with $|M_m^{\rm eff}| > 0.1 \,\mu_{\rm B}$,
since it is difficult to obtain $M_m^{\rm eff}$
accurately for when 
$\langle (M_m^z)^2\rangle < 0.01\,\mu_{\rm B}^2$.
Hence, 
Fig. 2 does not show the long-range tail of the magnetic-moment density 
away from the Fe site. 
It is cut-off when 
$|M({\bf r})| \ltsim 0.1\mu_{\rm B}$.

\bibliographystyle{apsrev4-1}

\bibliography{mhl2}

\end{document}